\documentclass[preprint,12pt]{elsarticle}

\usepackage{amssymb}
\usepackage{lipsum}
\usepackage{booktabs, multirow} 
\usepackage{soul}
\usepackage[table]{xcolor} 
\usepackage{changepage,threeparttable} 
\usepackage{acronym}
\usepackage[hidelinks,breaklinks]{hyperref}
\usepackage{caption}
\usepackage{subcaption}
\usepackage{amsmath}

\journal{Joule}

\begin{document}

\begin{frontmatter}



\title{Assessing the techno-economic benefits of LEMs for different grid topologies and prosumer shares}


\author[TUM]{Markus Doepfert\corref{cor1}}
\ead{markus.doepfert@tum.de}
\author[TUM]{Soner Candas}
\author[OTH]{Hermann Kraus}
\author[TUM]{Peter Tzscheutschler}
\author[TUM]{Thomas Hamacher}
\ead{thomas.hamacher@tum.de}

\cortext[cor1]{Corresponding author}

\affiliation[TUM]{organization={Technical University of Munich},
            addressline={Lichtenbergstrasse 4a}, 
            city={Garching},
            postcode={85748}, 
            state={Bavaria},
            country={Germany}}

\affiliation[OTH]{organization={Regensburg University of Applied Sciences},
            addressline={Seybothstrasse 2}, 
            city={Regensburg},
            postcode={93053}, 
            state={Bavaria},
            country={Germany}}

\begin{abstract}
    The shift towards decentralized and renewable energy sources has introduced significant challenges to traditional power systems, necessitating innovative market designs. Local energy markets present a viable solution for integrating distributed energy resources such as photovoltaic systems, electric vehicles, and heat pumps within various grid topologies. This study investigates the techno-economic benefits of local energy markets compared to conventional market designs, focusing on their impact on average energy prices and operational peak power, using a self-developed agent-based energy system simulation tool. Through comprehensive simulations across the countryside, rural, suburban, and urban grid topologies with varying penetration levels of the distributed energy resources, totaling 400 simulation setups, we demonstrate that local energy markets can enhance economic efficiency and grid stability with 99\,\% of the scenarios boasting lower average energy prices and 80\,\% lower operational peak power levels. Our findings suggest that local energy markets can play a role in the future energy system, especially in areas with high shares of PV and HP, provided that additional infrastructure, management costs, and bureaucratic complexity are kept to a minimum.
\end{abstract}

\begin{graphicalabstract}
    \begin{figure}[h!]
        \centering
        \includegraphics[width=\textwidth]{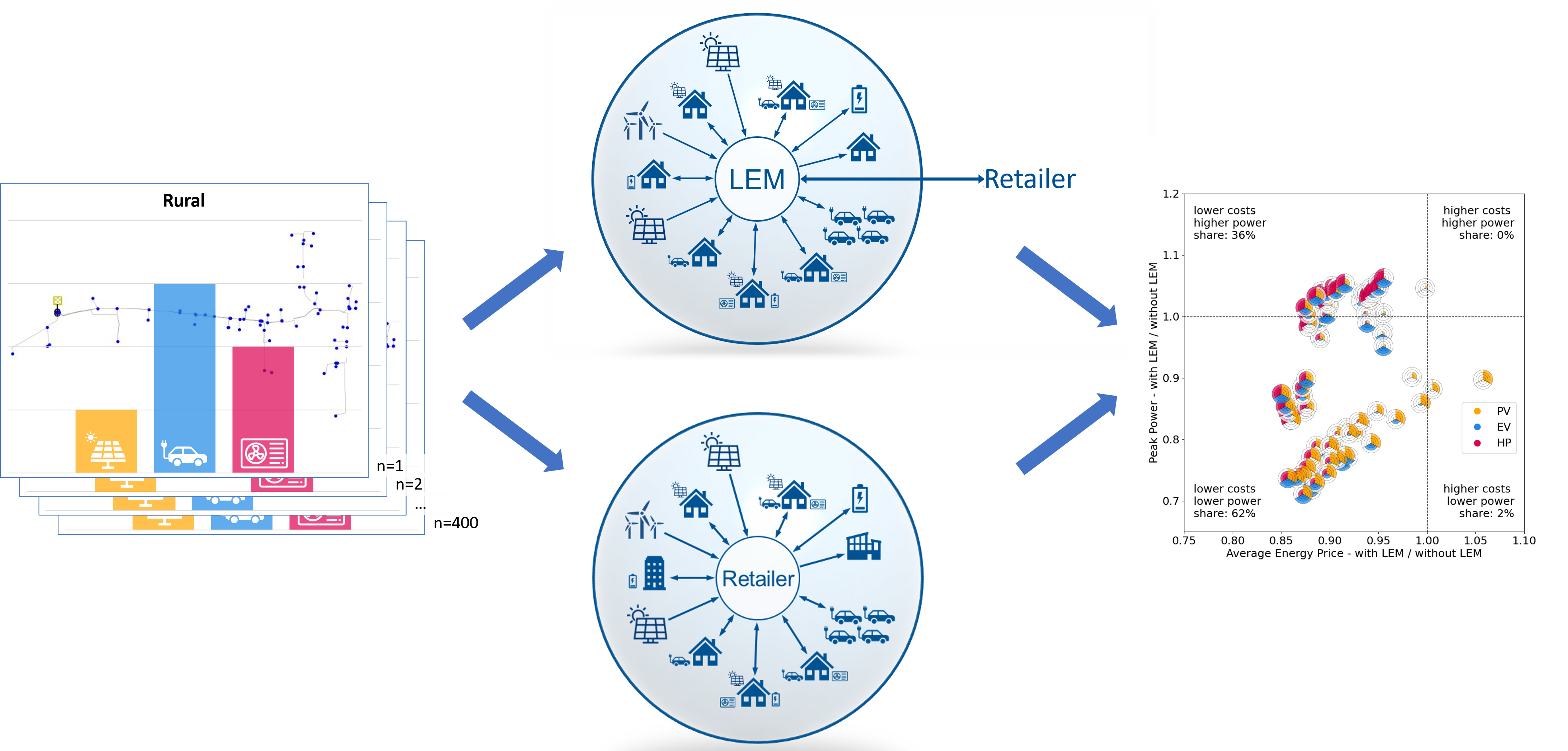}
        \caption{What economic and technical effect do local energy markets have on the energy system? To answer this question, we ran 400 different simulations with varying levels of PV, electric vehicles, heat pumps, and different grid topologies. Using our own agent-based simulation tool we ran the simulations with and without a local energy market. The results show that local energy markets can reduce economic and technical burdens in 99 and 80\,\%, respectively. Nevertheless, they cannot be introduced without any safeguarding measures as operation within technical constraints cannot be guaranteed by these types of markets alone.}
        \label{fig:grabs}
    \end{figure}
\end{graphicalabstract}

\begin{highlights}
    \item 400 different configurations of grid topology, PV, EVs and heat pumps are analyzed
    \item Local energy markets can reduce economic burden in 99\,\% of the cases and technical in 80\,\%
    \item They are especially suitable in areas where generation and demand coincide often
    \item Minimizing additional costs and bureaucracy is crucial for LEM viability
    \item Other mechanisms are required, however, to ensure technical stability 
\end{highlights}


\end{frontmatter}



\begin{abstract}
    The shift towards decentralized and renewable energy sources has introduced significant challenges to traditional power systems, necessitating innovative market designs. Local energy markets present a viable solution for integrating distributed energy resources such as photovoltaic systems, electric vehicles, and heat pumps within various grid topologies. This study investigates the techno-economic benefits of local energy markets compared to conventional market designs, focusing on their impact on average energy prices and operational peak power, using a self-developed agent-based energy system simulation tool. Through comprehensive simulations across the countryside, rural, suburban, and urban grid topologies with varying penetration levels of the distributed energy resources, totaling 400 simulation setups, we demonstrate that local energy markets can enhance economic efficiency and grid stability with 99\,\% of the scenarios boasting lower average energy prices and 80\,\% lower operational peak power levels. Our findings suggest that local energy markets can play a role in the future energy system, especially in areas with high shares of PV and HP, provided that additional infrastructure, management costs, and bureaucratic complexity are kept to a minimum.
\end{abstract}

\section{Introduction}

    The traditional model of electricity trading in Europe has long been predicated on the assumption that energy can flow freely without constraints. Historically, this assumption held mostly true as centralized power plants supplied electricity in a unidirectional, top-down manner. However, the energy landscape is undergoing a profound transformation, with decentralized renewable sources like wind and solar gradually supplanting centralized fossil fuel power generation -- a trend expected to persist~\cite{Gielen2019}. This shift comes at an economic cost. For instance, Germany has witnessed a steep increase in redispatch costs, soaring from 113 million euros in 2013 to 2.7 billion euros in 2022 as energy trades realized on the energy markets cannot be carried out physically due to grid congestions~\cite{BDEW2023}. Consequently, both the centralized nature of energy generation and the one-way flow of electricity are becoming outdated paradigms.
    
    Furthermore, alongside changes in electricity generation, a new type of energy system participant is emerging. In the past, households were conventional consumers characterized by predictable load patterns, easily forecasted using standard load profiles. However, with the adoption of photovoltaic (PV) systems, households are transitioning into prosumers -- consumers who also generate energy -- introducing a markedly different demand pattern. Moreover, as households incorporate battery storage, heat pumps (HPs), and electric vehicles (EVs), they evolve into flexumers. This shift grants them the ability to not only consume and produce electricity but also flexibly adjust their energy demand and generation which can reduce the burden on the grid~\cite{Needell2023}. Consequently, forecasting energy consumption becomes more challenging, and the proliferation of these electrical components adds volatility to the system. The existing market design inadequately reflects these technical realities, resulting in increased peak power loads on the grid partly due to the underincentivized utilization of available flexibility~\cite{Mugnini2023, Loschan2023}.
    
    Addressing these challenges necessitates a rethinking of the market design to align with the evolving technical realities~\cite{Qin2023}. A promising solution that has gained traction, particularly in Europe, is the incorporation of local energy markets (LEMs)~\cite{EU2018, EU2019}. LEMs, typically situated at the low-voltage level, facilitate energy trading among small-scale consumers and producers. The core idea is to provide price signals at the relevant technical grid level, motivating stakeholders to adjust their energy consumption and production patterns according to energy availability. As stakeholders actively balance generation and consumption at the local level, the expectation is that system costs will decrease as compliance with grid requirements improves~\cite{Khorasany2018}. In this context, the role of the wholesale market shrinks to that of a backup market, intervening only during energy deficits or surpluses, while the LEM operates independently within its region.
        
    While a considerable body of scientific literature has delved into the intricate design and real-world implementation of LEMs, certain critical aspects have garnered particular attention. One such facet is the choice between decentralized and centralized trading mechanisms~\cite{Zaidi2018, Hussain2018, Gazafroudi2021, Ali2023}. Decentralized or peer-to-peer (P2P) trading involves direct transactions among participants without central oversight, often coupled with blockchain technology. In contrast, centralized markets provide a platform where participants trade, with an entity ensuring system stability~\cite{Khorasany2018}. 
    Another design consideration in LEM design is the clearing mechanism, encompassing decisions regarding when and how the market should be cleared, such as ex-ante or ex-post clearing, and day-ahead or intraday clearing~\cite{Kim2023}.
        
    Beyond the question of market design, real-world implementations of LEMs have been the focus of numerous pilot projects aimed at testing various integration approaches into existing grids~\cite{Bjarghov2021}. These initiatives have been executed in diverse regions, including the USA~\cite{Mengelkamp2018}, Germany~\cite{Vasconcelos2019}, and Switzerland~\cite{Worner2019}.
        
    In this paper, however, we want to step back from the detail discussions and examine the broader picture. Existing studies often assume a fixed setup for local energy markets. Still, the composition of components (i.e., PV/battery systems, EVs, HPs) and the grid topology can vary significantly from one region to another. Therefore, it is imperative to not only scrutinize the specifics of the LEM design and implementation but also discern the circumstances under which LEM deployment becomes economically and technically sensible.
    
\begin{figure}
        \centering
        \includegraphics[width=.8\textwidth]{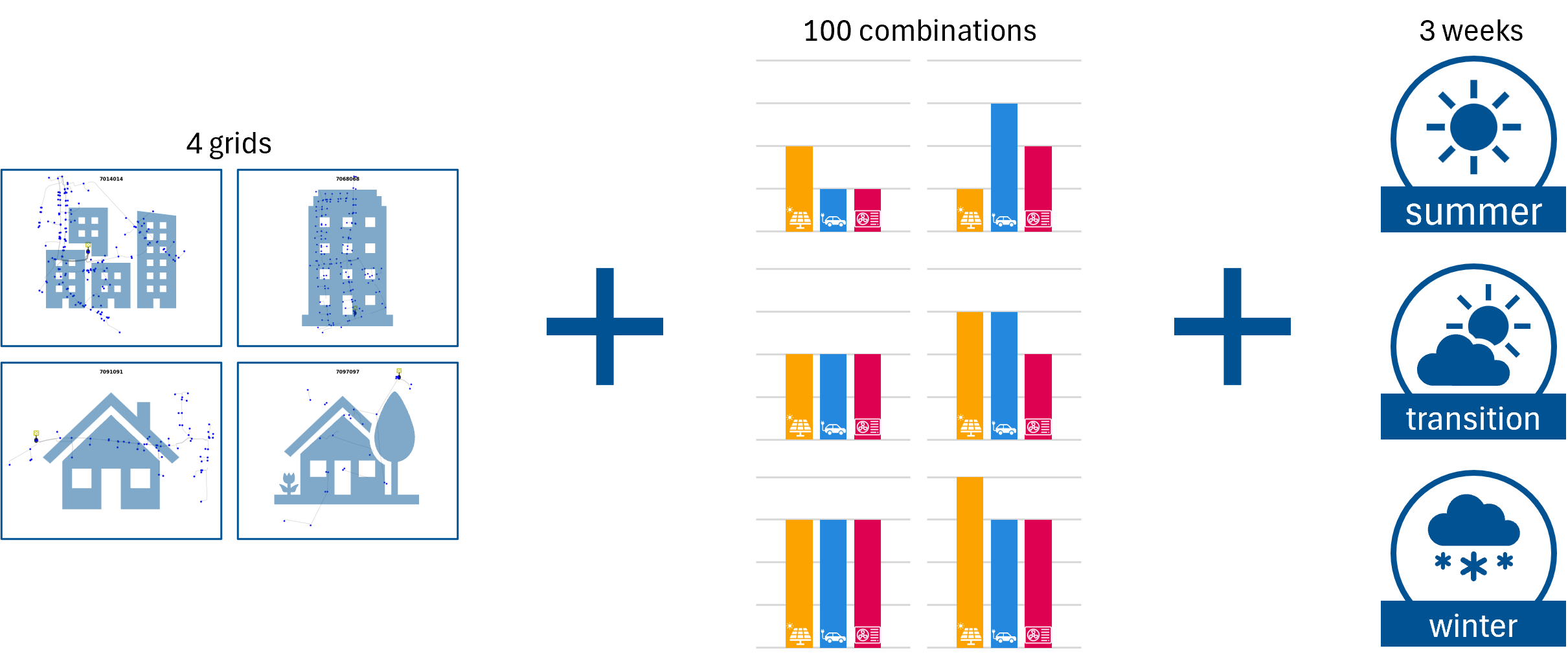}
        \caption{Methodology for the creation of the scenarios to be examined}
        \label{fig:scenario_creation}
    \end{figure}

    To address this overarching question, our research undertakes a comprehensive analysis of LEMs across various grid topologies and different penetration levels of energy devices. We consider four distinct grid topologies, each representing typical countryside, rural, suburban, and urban areas. Additionally, we vary the proportion of PV/battery systems, EVs, and HPs in increments of 25\,\%, resulting in a total of 500 distinct scenarios, as shown in \autoref{fig:scenario_creation}. All 100 scenarios that do not have any PV have been disregarded, however, as without generation will be no price formation on the market. Each of the remaining 400 scenarios undergoes a comprehensive analysis over three representative weeks, covering the seasons of summer, winter, and spring/autumn, i.e. the transition periods. For further information about the simulation setup, we refer you to \autoref{ssec:methodology}. Our analysis primarily focuses on two key dimensions: the economic impact of LEMs, as assessed by changes in average energy prices (AEPs) compared to scenarios without LEMs, and the influence of LEMs on operational peak power (OPP), as assessed by changes in the peak 15\,\% power flows at the transformer compared to scenarios without LEMs. We chose these two key parameters as they offer good insights into the economic and technical impact during the operation of the energy system. By assessing how LEMs influence these two key parameters for a range of different system setups, we aim to provide insights into when and to what extent LEMs can play a role in easing economical and technical burdens on the system for other researchers as well as policy decision makers.

\section{Results}

    The results section analyze how LEMs influence the average energy price (AEP) and operational peak power (OPP) across various grid topologies and technology penetration rates, namely photovoltaics with batteries (PV), electric vehicles (EVs), and heat pumps (HPs).
    
    
    The analysis proceeds as follows: \autoref{ssec:results_technical_economical} presents a comparative overview of the relative AEP and OPP across all relevant scenarios, aiming to identify broad trends and specific conditions under which LEMs yield the most significant benefits in comparison to market setups without LEMs. 
    Subsequently, we focus on the rural and suburban results for our further results as these represent the countryside and urban topology, respectively, rather well. Any observations where the respective topologies are not aligned will be addressed in the respective section of this paper.
    In \autoref{ssec:results_economic} and \autoref{ssec:results_technical}, we delve deeper into the economic and technical dimensions, respectively, examining the absolute impacts of these variables under different technology penetrations. In \autoref{ssec:results_sensitivity} we conduct an in-depth analysis of the influence of each technology -- PV, EV, and HP -- on the overall energy system from both economic and technical perspectives to reveal how their shares influence the energy system.

    The AEP is calculated considering both the costs of obtained energy as well as the lost earnings from the self-consumed PV generation: 
        
		\begin{minipage}{\textwidth}
            \begin{equation}
                p_\mathrm{AEP} = \frac{\sum_{t=0}^{n} E_{t,\mathrm{self}} * p_\mathrm{PV} + E_{t,\mathrm{market}} * (p_{t,\mathrm{market}} + p_\mathrm{levies})}{\sum_{t=0}^{n} E_{t,\mathrm{self}} + E_{t,\mathrm{market}}}
                \label{eq:cost_avg}
            \end{equation}
			\flushleft
            \footnotesize{
    			\begin{align*} 
    				\text{where}\\
    				E_{t,\mathrm{self}}	    &= \text{self-consumed energy at time $t$}	            & \left[E_{t,\mathrm{self}}\right]      &= \mathrm{kWh}			    \\
    				p_\mathrm{PV}   	    &= \text{feed-in tariff} = 0.0827	                    & \left[p_\mathrm{PV}\right]            &= \mathrm{\mathrm{\text{€}/kWh}}  \\
    				E_{t,\mathrm{market}}   &= \text{energy bought from the market at time $t$}     & \left[E_{t,\mathrm{market}}\right]    &= \mathrm{kWh}	            \\
    				p_{t,\mathrm{market}} 	&= \text{market price per unit of energy at time $t$}   & \left[p_{t,\mathrm{market}}\right]    &= \mathrm{\mathrm{\text{€}/kWh}}  \\
    				p_\mathrm{levies}		&= \text{grid-fees and levies per unit of energy}       & \left[p_\mathrm{levies}\right]        &= \mathrm{\mathrm{\text{€}/kWh}}	\\
    				n 		                &= \text{total number of time intervals.}				&										&						    \\
    			\end{align*} 
            }
        \end{minipage}
 
        The OPP is calculated using the highest 15\,\% of the absolute power flow values over the transformer to represent the load the transformer is under during operation. The average of the top 15\,\% peak power $\bar{P}_{\mathrm{OPP}}$ values is calculated by:

        \begin{minipage}{\textwidth}
            \begin{equation}
                \bar{P}_{\mathrm{OPP}} = \frac{1}{k} \sum_{i=1}^{k} P_{\text{sorted},i}
                \label{eq:power_peak}
            \end{equation}
            \flushleft
            \footnotesize{
                \begin{align*}
                    \text{where}\\
                    k                       &= \text{number of time steps considered for the top 15\,\% peak power calculation}             &                           & \\
                    P_{\text{sorted},i}    &= \text{the $i^\mathrm{th}$ power value in the sorted list of absolute power flow values}   & [P_{\text{sorted},i}]     &=  \mathrm{kW}.\\
                \end{align*}
            }
        \end{minipage}
        
    \subsection{Relative impact of LEMs  on the average energy price and operational peak power}\label{ssec:results_technical_economical}
        
\begin{figure}
            \centering
            \begin{subfigure}[b]{0.49\textwidth}
                \centering
                \includegraphics[width=\textwidth]{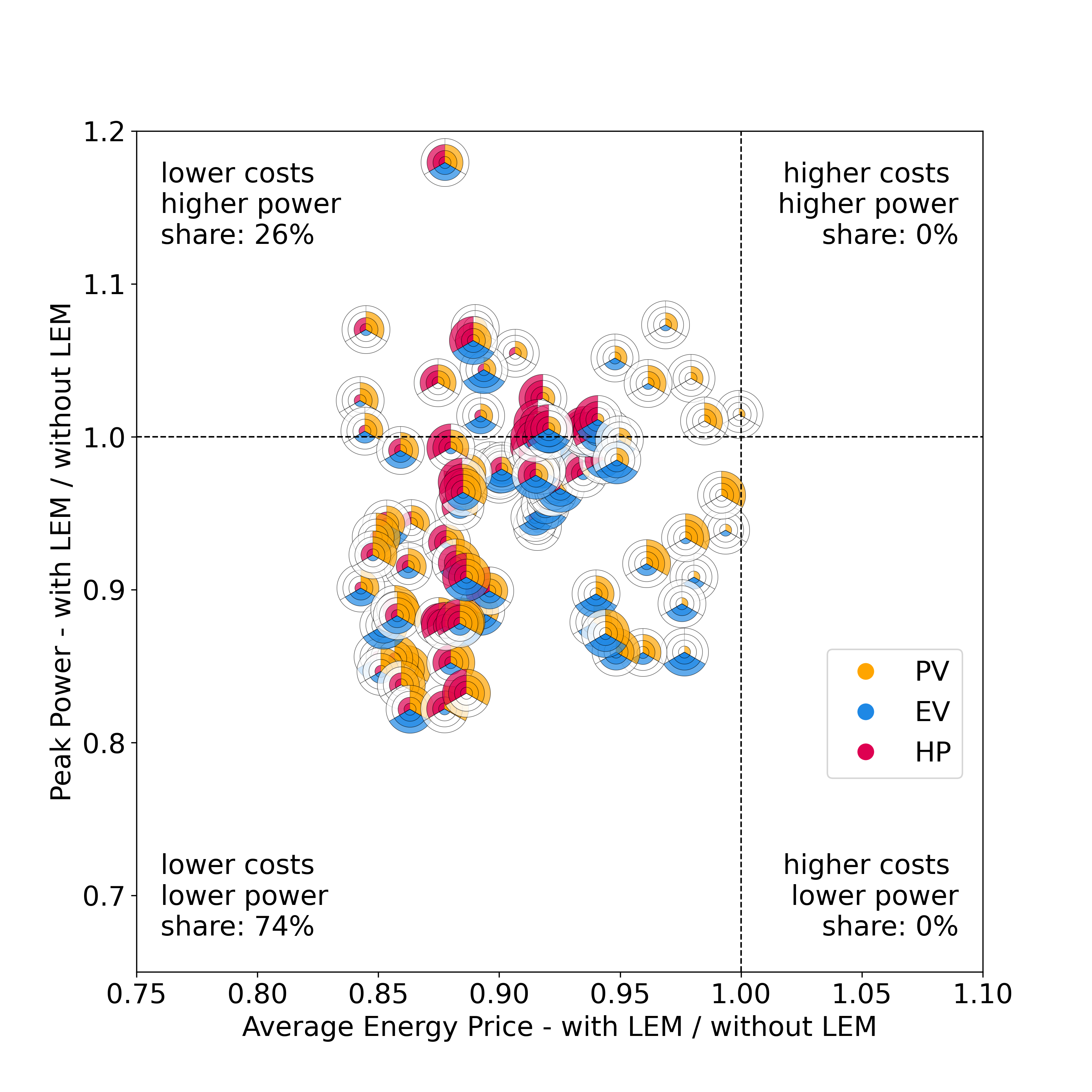}
                \caption{Countryside}
                \label{fig:price_power_countryside}
            \end{subfigure}
            \hfill
            \begin{subfigure}[b]{0.49\textwidth}
                \centering
                \includegraphics[width=\textwidth]{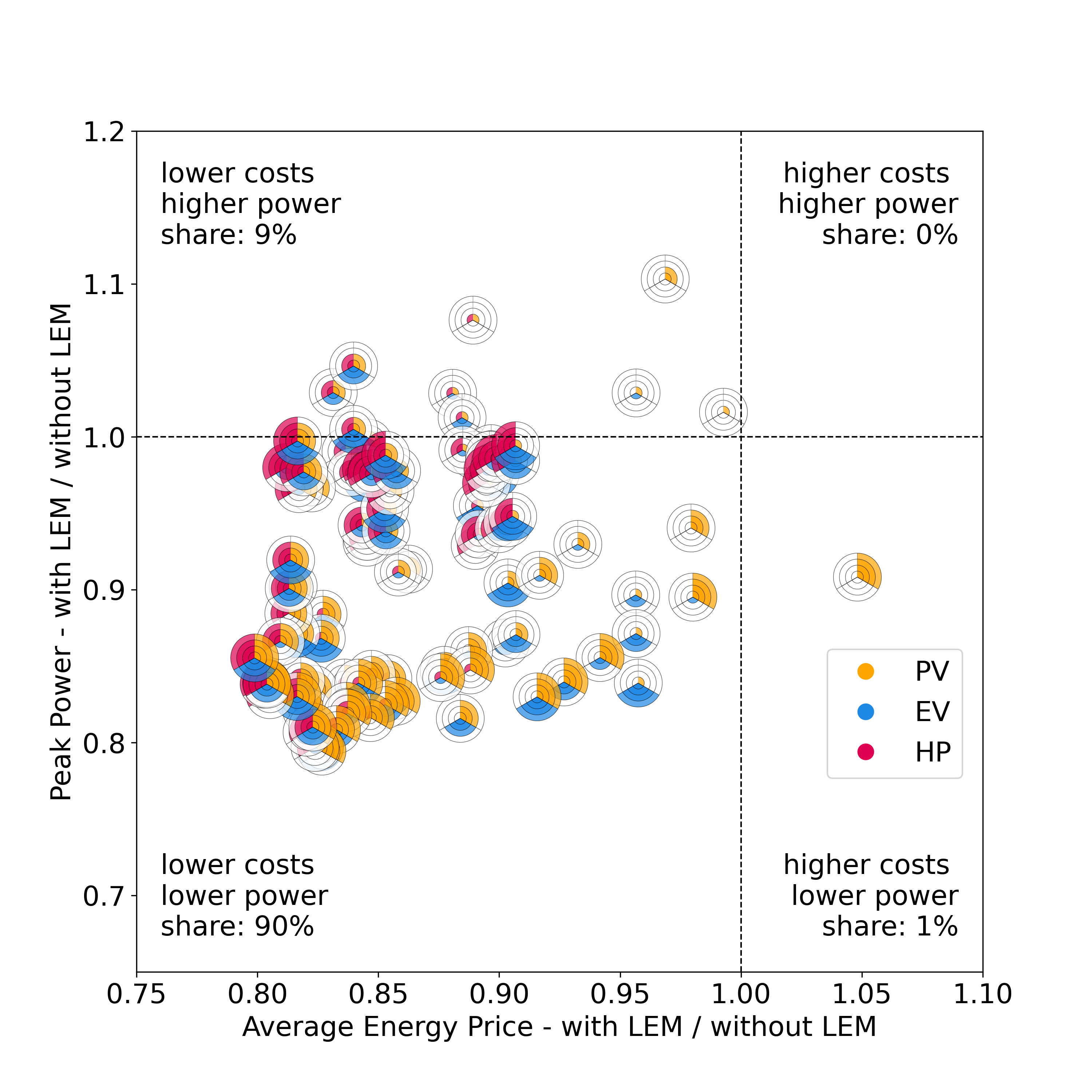}
                \caption{Rural}
                \label{fig:price_power_rural}
            \end{subfigure}
        
            \vspace{1ex} 
        
            \begin{subfigure}[b]{0.49\textwidth}
                \centering
                \includegraphics[width=\textwidth]{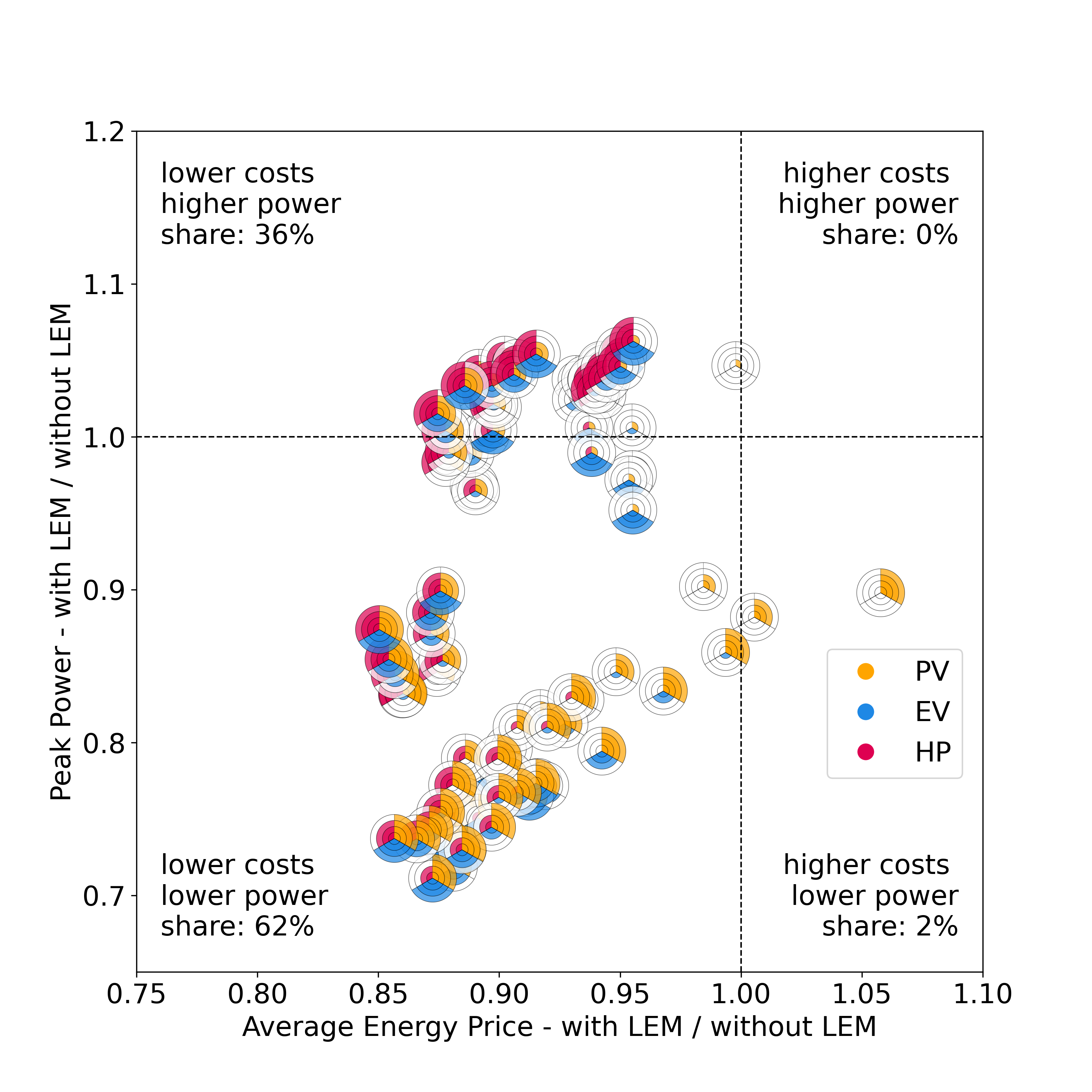}
                \caption{Suburban}
                \label{fig:price_power_suburban}
            \end{subfigure}
            \hfill
            \begin{subfigure}[b]{0.49\textwidth}
                \centering
                \includegraphics[width=\textwidth]{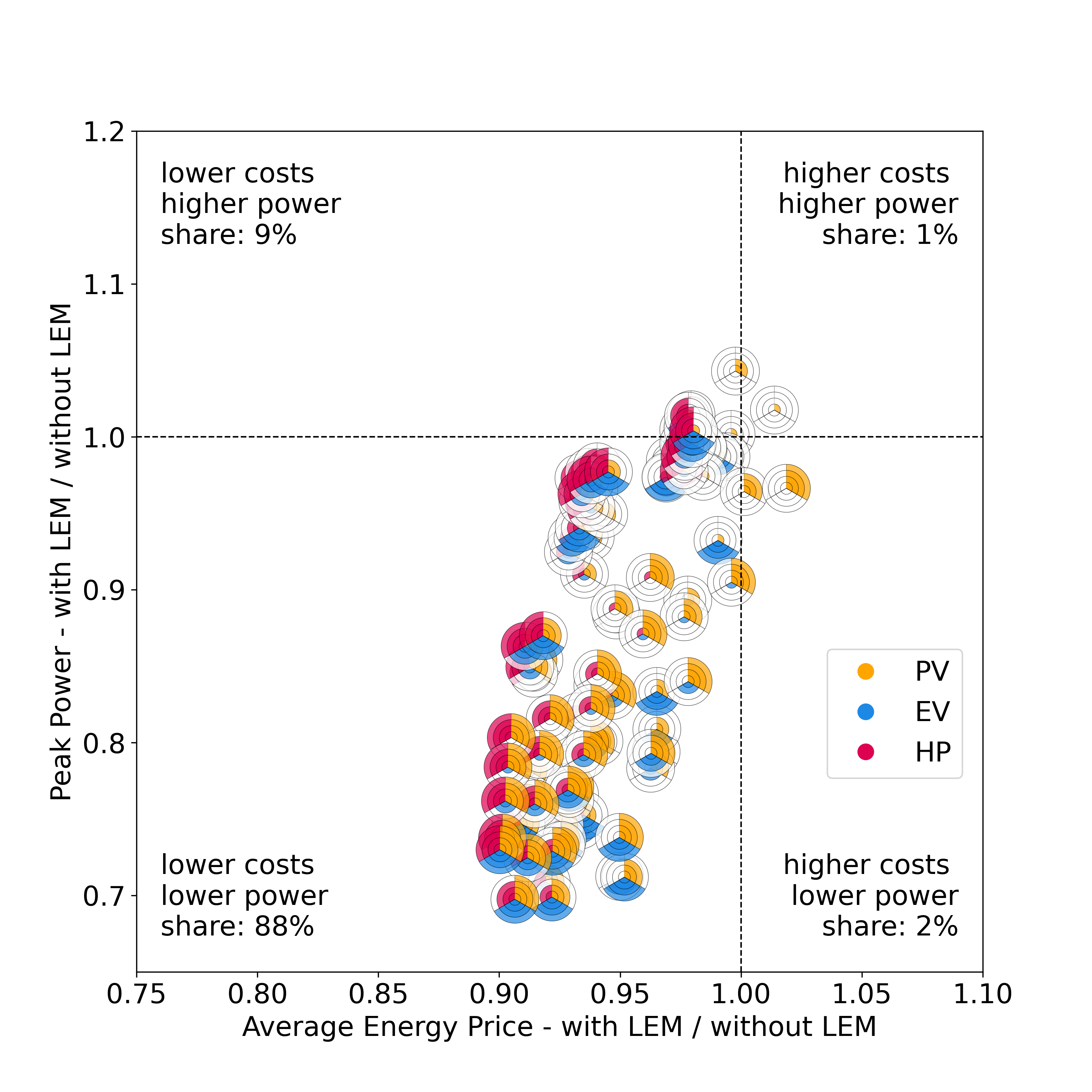}
                \caption{Urban}
                \label{fig:price_power_urban}
            \end{subfigure}
        
            \caption{Comparison of the average energy price and operational peak power ratio for scenarios with and without LEMs for every combination of PV, EV and HP}
            \label{fig:price_power}
        \end{figure}

        We first compare the relative impact of Local Energy Markets (LEMs) on economic and technical aspects across four different grid topologies, as depicted in \autoref{fig:price_power}. Each subfigure is divided into four quadrants, indicating whether scenarios with LEMs result in higher or lower costs and power levels. Across all grid topologies, the majority of scenarios demonstrate benefits in both aspects. However, the extent of these benefits varies significantly between topologies. Notably, grid topologies with fewer participants, such as countryside and rural grids, exhibit more variability compared to suburban and urban grids, where the impact of individual participants is less pronounced, reducing the influence of each individual and outliers.

        In the countryside grid, no scenario falls into the higher-costs/higher-power or higher-costs/lower-power quadrants. 
        The relative energy price ratios range from 0.84 to 1.00, indicating that systems with LEMs can be up to 16\,\% cheaper. Scenarios lacking HPs tend to show comparatively higher costs albeit still being below those without a LEM. The introduction of more flexibility generally shifts the scenarios to lower costs. 
        The impact of LEMs on power levels is less straightforward. While 74\,\% of scenarios show lower operational peak power (OPP), 26\,\% show higher power levels, with ratios ranging from 0.82 to 1.18. This variability suggests that outliers in low-participant grids have a significant impact, unlike in larger grids where the effect is averaged out. Overall, 74\,\% of the scenarios in the countryside grid fall within the 0.9 to 1.1 ratio band, indicating a relatively small impact of LEMs on OPP.
        
        Similarly, the rural grid has no scenarios in the higher-costs/higher-power quadrant. 
        However, the 100\,\% PV scenario is more costly with the chosen LEM design. The ratios of the AEP range between 0.80 and 1.05. Increased flexibility, in the form of EVs and especially HPs, lowers costs compared to scenarios without LEMs. 
        Regarding OPP, 91\,\% of scenarios exhibit lower power levels, with ratios between 0.79 and 1.10. Scenarios with high PV shares ($\geq75\,\%$) tend to have lower OPPs, with 78\,\% falling below 0.9. Conversely, scenarios with high HP shares ($\geq75\,\%$) tend to have higher OPPs, with 70\,\% exceeding 0.9.
        
        The suburban topology shows a more organized pattern. The scenarios can be grouped into three clusters. The first cluster centers around the 1-ratio of OPP, with most scenarios having higher HP shares than PV shares, encompassing 45\,\% of the scenarios. The second cluster, which is dominated by high shares of PV and low shares of HPs, runs diagonally from the higher-costs/lower-power to the lower-costs/lower-power quadrant, including another 45\,\% of scenarios. The smallest cluster, comprising 10\,\% of scenarios, is around the 0.85-cost and 0.85-power ratio and contains scenarios with high shares of both PV and HP. 
        Overall, the cost ratio varies between 0.85 and 1.06, with 98\,\% of scenarios exhibiting lower costs. Similar to the rural grid, cost outliers are primarily scenarios with only PV. As the share of flexible components increases, costs decrease. 
        The OPP ratio ranges from 0.71 to 1.06, with 36\,\% of scenarios showing higher OPP ratios, primarily those with 25\,\% PV or equal/greater HP shares compared to PV. Conversely, the 64 scenarios with lower OPP ratios often have higher PV shares and HP shares that are lower or equal to that of PV.
        
        The urban topology displays the most organized pattern. 
        The average energy price ratio ranges from 0.90 to 1.02, with 97\,\% of scenarios showing ratios $\leq 1$. All scenarios where LEMs are more expensive lack HPs and EVs. As the share of these technologies increases, the ratio tilts in favor of LEMs. 
        Additionally, 90\,\% of scenarios exhibit lower OPP levels with LEMs, with ratios between 0.70 and 1.04. Similarly to the AEP there is a noticeable tendency that higher shares of PV generally correlate with lower OPP ratios. Specifically, scenarios with PV shares $\geq75\,\%$ have an average OPP ratio of 0.79, whereas scenarios with shares $\leq50\,\%$ average at 0.95.

    \subsection{Absolute impact of LEMs on the average energy price}\label{ssec:results_economic}
    
\begin{figure}
            \centering
            \begin{subfigure}[b]{0.49\textwidth}
                \centering
                \includegraphics[width=\textwidth]{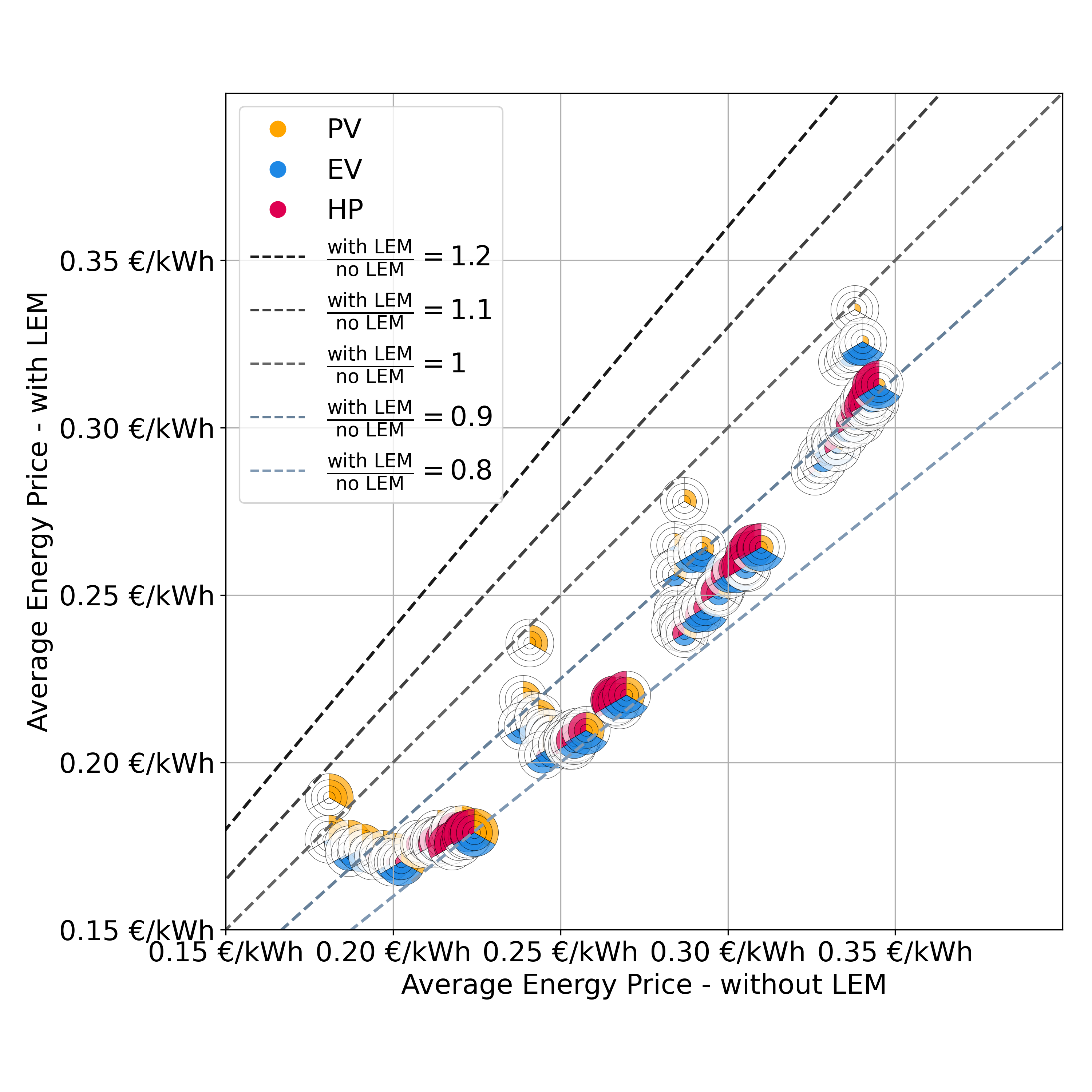}
                \caption{Rural}
                \label{fig:price_rural}
            \end{subfigure}
            \hfill
            \begin{subfigure}[b]{0.49\textwidth}
                \centering
                \includegraphics[width=\textwidth]{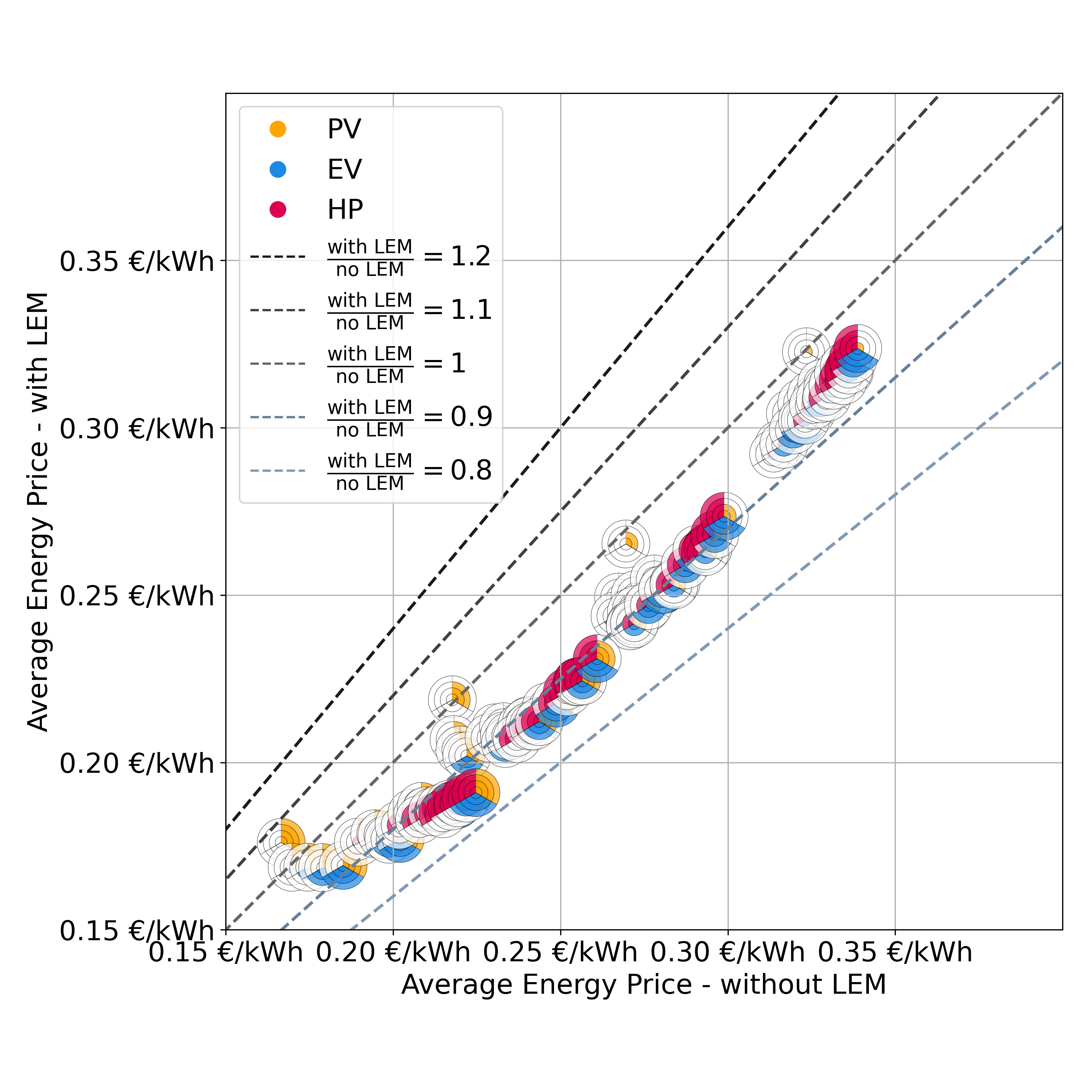}
                \caption{Suburban}
                \label{fig:price_suburban}
            \end{subfigure}
        
            \caption{Comparison of the absolute energy prices for scenarios with and without LEMs for every combination of PV, EV and HP}
            \label{fig:price}
        \end{figure}

        Ratios serve as a good first indicator to assess the LEMs impact. However, we also need to analyze the absolute economic impact of LEMs as some of the information is lost using ratios. For this reason we analyze the absolute AEPs with and without LEMs in this section. The results are shown in \autoref{fig:price} for the rural and suburban topology.  It plots the absolute AEP with and without LEM with dashed lines indicating the ratio between the two values.

        Both topologies exhibit a similar pattern. The results can be divided into four clusters, each representing a different share of PV systems. Generally, the AEP decreases as the PV share increases, regardless of the presence of an LEM. This trend is due to the lower energy costs of self-consumed PV energy compared to grid-supplied energy. 
        In the rural grid, the absolute costs for scenarios with LEMs range from 0.170\,€/kWh to 0.335\,€/kWh, and from 0.181\,€/kWh to 0.345\,€/kWh without LEMs. In the suburban grid, the values range from 0.169\,€/kWh to 0.324\,€/kWh and 0.167\,€/kWh to 0.339\,€/kWh respectively.
        
        Each cluster follows a similar pattern where, in scenarios with only PV, the impact of LEMs on prices is minimal. However, as more flexibility is introduced, the prices shift in favor of LEMs, as participants can better utilize both their own energy and the available energy on the market. Consequently, scenarios with shares of HP and EV of $\geq50\,\%$ have an average ratio of 0.85 across both topologies, while those with $<50\,\%$ have an average ratio of 0.94. 
        Nonetheless, the absolute costs vary significantly in these scenarios. Scenarios with the highest HP-to-PV ratios also observe the highest absolute prices. This is because using an HP for heating significantly increases the electricity demand. On average, households in Germany have an annual electricity demand of 2,900\,kWh \cite{GermanFederalStatisticsOffice2021}, while a heat pump adds another 5,000\,kWh, nearly tripling the total electricity demand  \cite{Schlemminger2022}. Since most of this demand occurs during the winter months, when PV generation is minimal, the additional electricity must be purchased from the retailer at higher prices.
    
    \subsection{Absolute impact of LEMs on the operational peak power}\label{ssec:results_technical}

\begin{figure}
            \centering
            \begin{subfigure}[b]{0.49\textwidth}
                \centering
                \includegraphics[width=\textwidth]{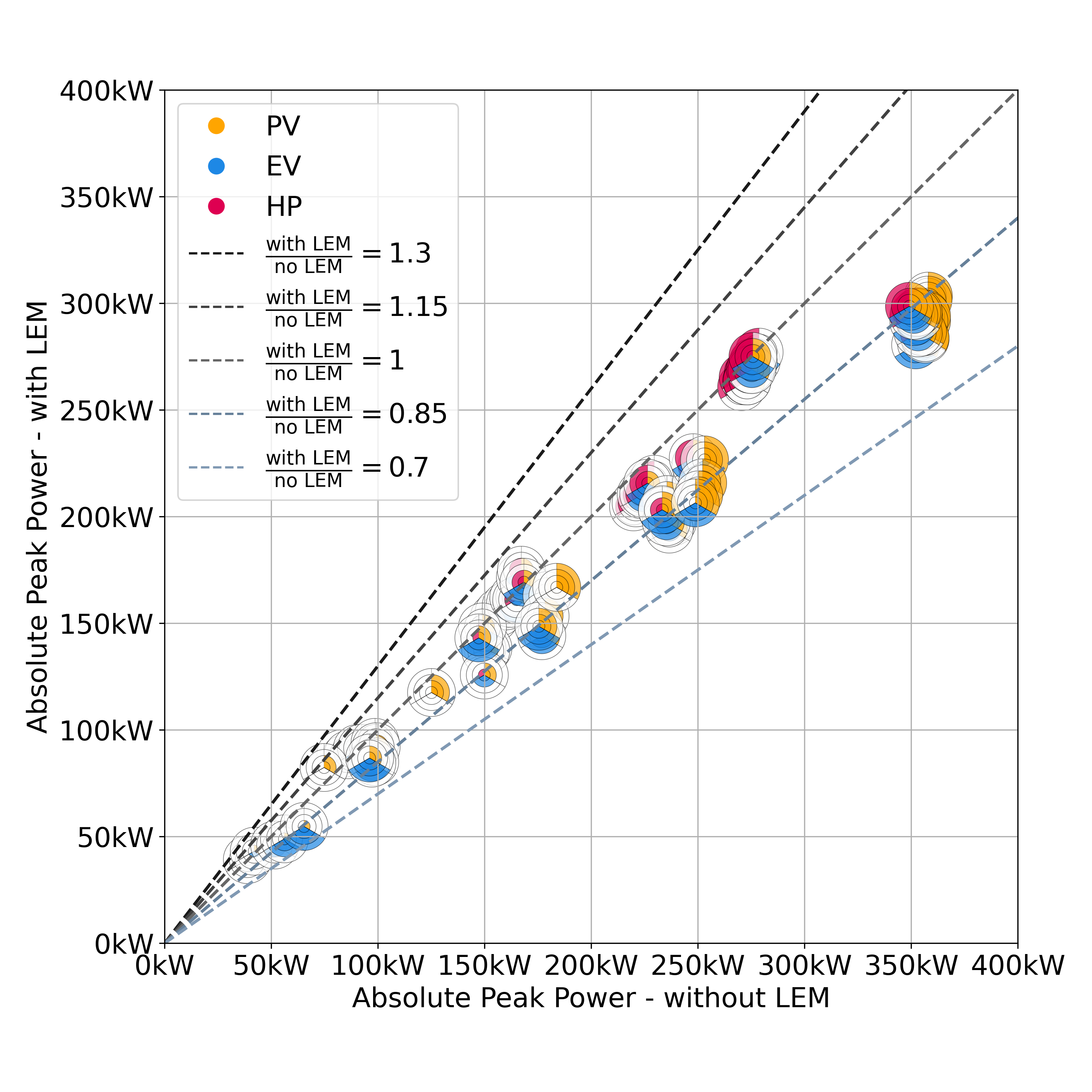}
                \caption{Rural}
                \label{fig:power_rural}
            \end{subfigure}
            \hfill
            \begin{subfigure}[b]{0.49\textwidth}
                \centering
                \includegraphics[width=\textwidth]{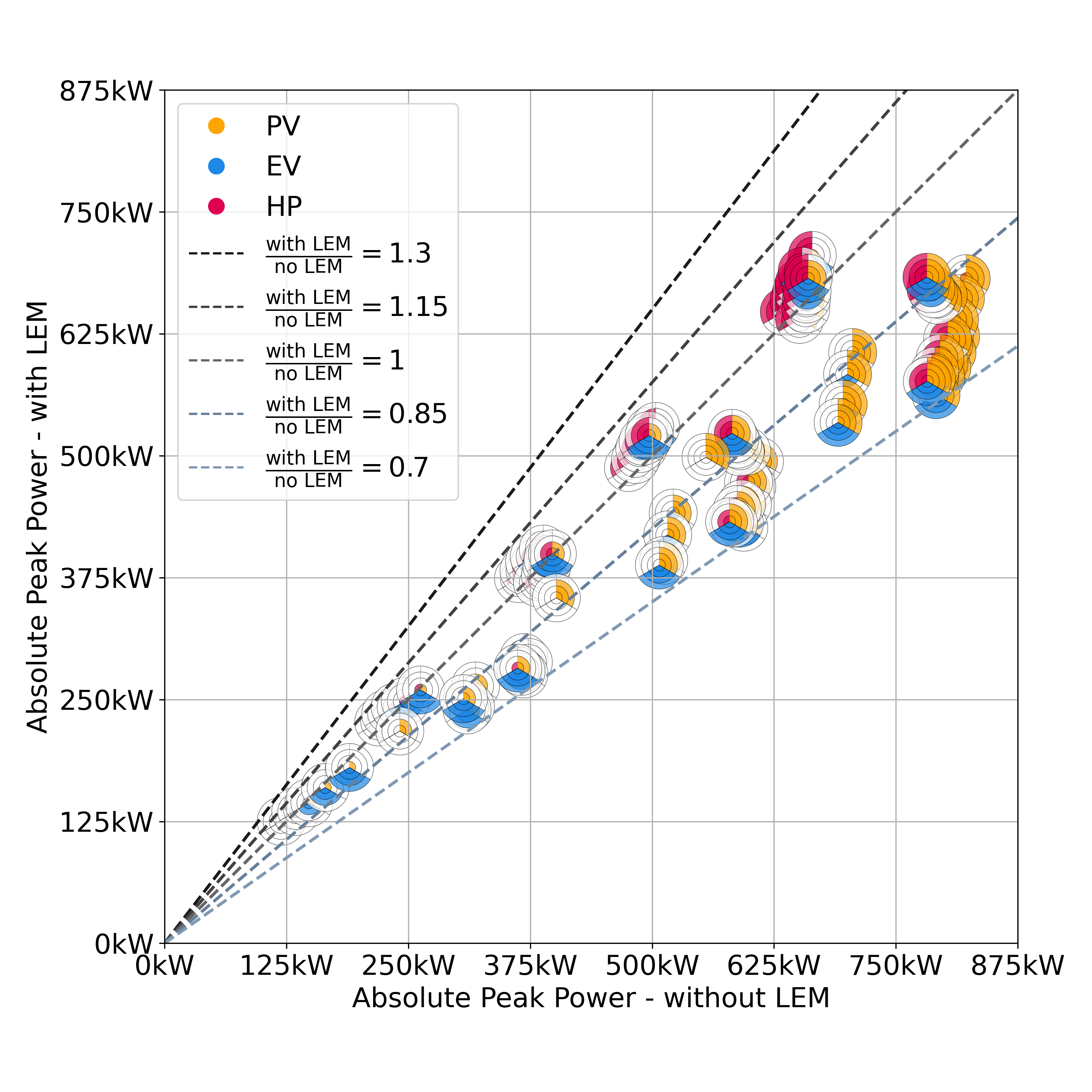}
                \caption{Suburban}
                \label{fig:power_suburban}
            \end{subfigure}
        
            \caption{Comparison of the operational peak power values for scenarios with and without LEMs for every combination of PV, EV and HP}
            \label{fig:power}
        \end{figure}

        The patterns observed \autoref{fig:power} are less clear-cut than those for AEP shown in \autoref{ssec:results_economic}. As expected, OPPs increase with the increase in the components' shares, which holds true for all topologies. Furthermore, it can be observed that the scenarios can mostly be clustered into PV or HP dominant groups with EVs playing a less dominant role as no EV dominant cluster can be observed. Instead, it is either the PV capacity whose generation in summer determines the OPP or the HP capacity with its demand in winter. Scenarios that are either PV or HP dominant remain so in both market setups regardless if there is a LEM or not as LEMs only reduce the OPPs proportionally while the general pattern of either summer- or winter-dominant scenarios remains. Consequently, the scenarios can be clustered by their most dominant technology.

        In the rural topology, the lowest OPP values, with and without LEM, are both 39\,kW, while the highest values are 303\,kW and 358\,kW, respectively. Scenarios, where the PV share exceeds the HP share, are consistently PV dominant, whereas those with higher HP shares are HP dominant. When the shares are equal, it can lean either way, with 25/25\,\%, and 50/50\,\% being PV dominant, and 75/75\,\%, and 100/100\,\% being HP dominant. Specifically, of the 20 combinations with equal PV and HP shares, 11 are PV dominant, and 9 are HP dominant. In most scenarios, the EV share has a negligible influence, except in scenarios with low shares of PV and HP, such as the 25/0\,\% scenarios, where a notable influence is observed. In these scenarios, OPPs increase, but those with LEMs utilize additional flexibility and have lower OPPs. In most other cases, EVs have a neutral or slightly positive effect. The line of best fit shows a slope of 0.88 with the coefficient of determination $R^2=0.99$, indicating that LEMs have an increasingly positive effect as the shares of PV, HPs, and EVs increase. These scenarios are the most critical for grid planners to consider when deploying reinforcements to manage power thresholds.
        
        Similar trends are observed in the suburban topology, where OPP values range from 125/120\,kW to 706/822\,kW (with/without LEM). The patterns are similar but more dispersed. Similar to the rural topology, scenarios with a higher PV than HP share have OPP values defined by the summer months while the reverse is true for scenarios with higher HP than PV shares. However, in this case, all scenarios with a parity of PV and HP shares were defined by the heating demand in winter. We attribute this fact to the lower ratio of PV production to HP demand as the roof area is smaller in suburban areas while the surface area of the buildings per occupant is still relatively large as dwellings are still dominated by single-family homes. The slope of the best-fit curve lies at 0.94, with $R^2=0.99$ showing the same tendency albeit to a lesser extent.
        
        In contrast, the urban grid has less space for PV, however, due to multi-family homes making up the greatest share in the building stock, their specific heat demand is also significantly lower than in suburban areas. For this reason, scenarios with higher HP shares tend to show lower OPP ratios than in the suburban topology.
        Overall it can be observed that LEMs are more effective in reducing peak loads in suburban and urban areas, with average OPP ratios of 0.89 and 0.87, respectively, compared to 0.96 and 0.91 in countryside and rural grids.
        
    \subsection{Influence of different penetration levels of plants on the average energy price and operational peak power} \label{ssec:results_sensitivity}
    
\begin{figure}
            \centering
            \begin{subfigure}[b]{0.49\textwidth}
                \centering
                \includegraphics[width=\textwidth]{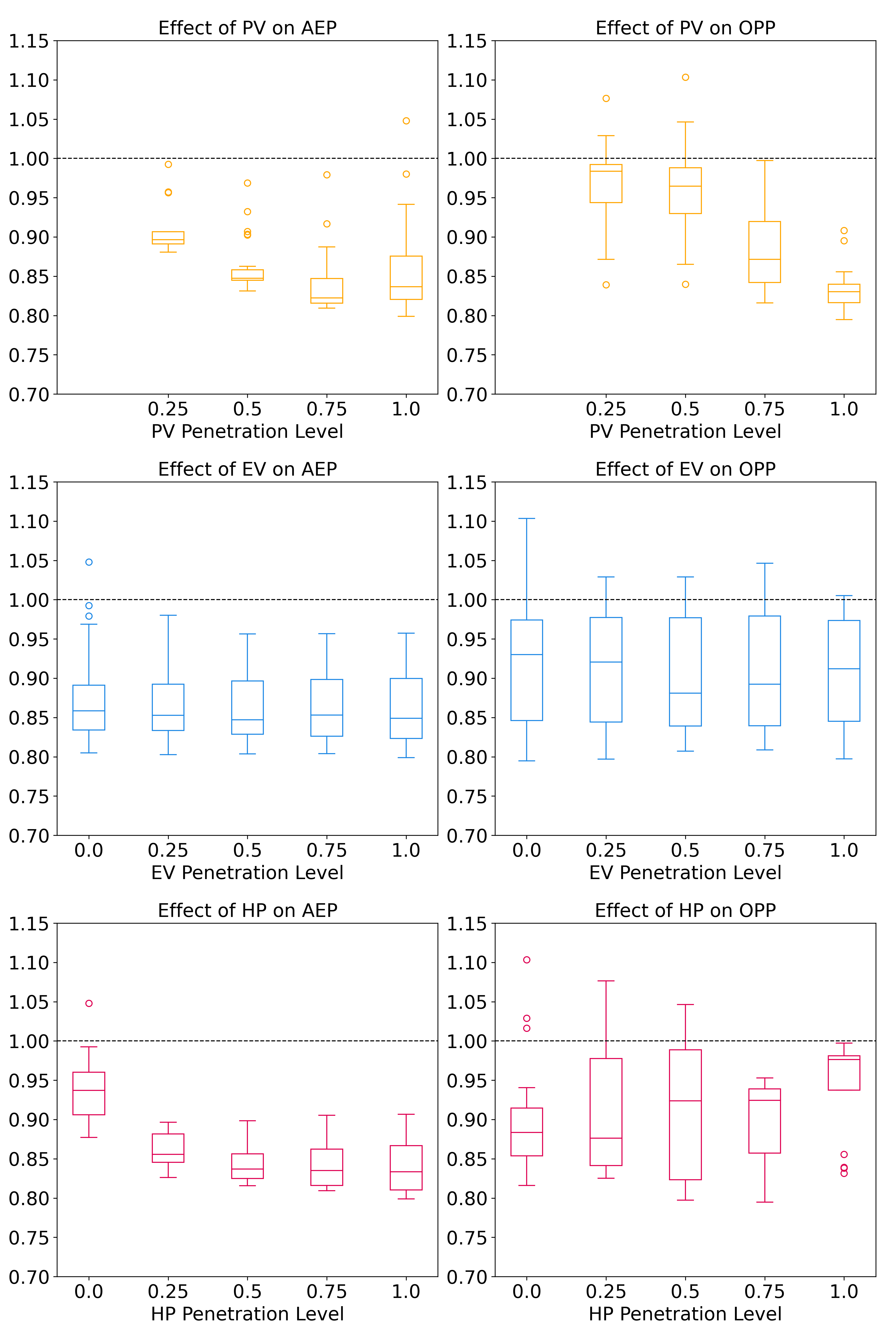}
                \caption{Rural}
                \label{fig:indepth_rural}
            \end{subfigure}
            \hfill
            \begin{subfigure}[b]{0.49\textwidth}
                \centering
                \includegraphics[width=\textwidth]{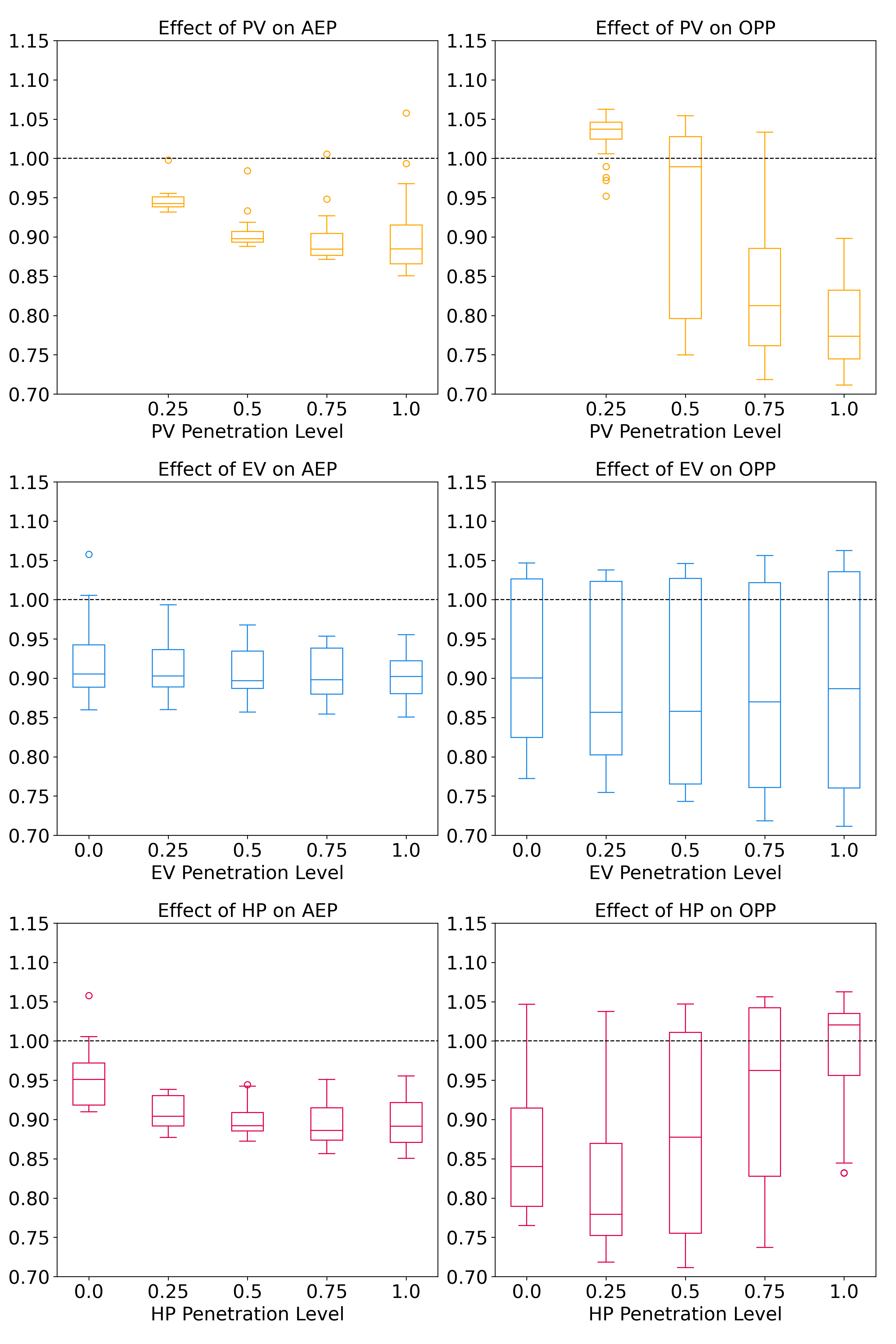}
                \caption{Suburban}
                \label{fig:indepth_suburban}
            \end{subfigure}
        
            \caption{In-depth analysis of the effects under specific penetration rates of PV, EV and HP on the average energy price and the operational peak power values in LEMs}
            \label{fig:indepth}
        \end{figure}

        So far, we have primarily focused on the impact of LEMs on the Average Energy Price (AEP) and the Operational Peak Power (OPP) levels. However, it is equally important to investigate how varying penetration levels of photovoltaics (PV), electric vehicles (EVs), and heat pumps (HPs) influence these outcomes across the scenarios. The interplay between these components significantly affects the stability, efficiency, and economic viability of energy systems. This section provides an in-depth analysis by comparing the penetration levels of each component against their impact on AEP and OPP, as shown in \autoref{fig:indepth} for the rural and suburban grids. Each boxplot illustrates the penetration levels of the components versus the AEP or OPP ratios for markets with and without LEM. The dashed line at $y=1$ represents equilibrium for both market designs.

        Starting with the influence of PV on the AEP, it can be observed that under most circumstances, a higher share of PV results in lower energy costs. For shares $\leq50\,\%$ the decrease can be confidently projected regardless of the topology. For greater shares, the costs tend to decrease further, albeit less strongly and with greater variance, settling at an average median value of 0.88 across all topologies. We attribute this increased variance to the high variations of available flexibilities in the 100\,\%-PV scenarios. Without sufficient flexibility resources the price can even turn out higher than without the LEM due to the additional costs from balancing energy.
        Similar to the costs, the relative OPP can be further reduced with higher shares of PV regardless of topology. While LEMs rarely have a strong impact on scenarios with $S_\mathrm{PV}\leq50\,\%$, they can have significant positive impacts in scenarios with higher shares. For $S_\mathrm{PV}>50\,\%$ median values are around 0.85 for the rural and 0.79 for the suburban grid (countryside: 0.92 and urban: 0.79). Thus, LEMs are especially beneficial in high load scenarios using the localized price incentive to effectively lower OPP levels.
        In conclusion, we derive that higher shares of PV have a generally positive effect on both the AEP as well as the relative OPP.
        
        Next, we investigate the EVs influence. Regarding the economic influence, their impact is positive albeit small. The median is almost constant across all penetration levels and topologies, hovering at about 0.86 for the rural and 0.91 for the suburban grid. However, the uncertainty of the price reduction decreases with an increased share of EVs.
        Regarding the relative OPP the results suggest that EVs have a small yet positive effect in LEMs. However, the uncertainty is high throughout all penetration rates.
        Overall, it can be observed that the EV's influence on both the economic and technical side is positive yet small.
        
        Lastly, we turn our attention to the heat pumps. In terms of economic effects, we observe a clear positive downwards trend that wears off for shares $\geq50\,\%$ settling at a median value of around 0.84 and 0.89 for the rural and suburban grid, respectively. The decrease in cost is due to the increased time when generation and demand can be matched, especially during the transition period where days with high PV yield are still numerous while there is also a significant amount of heating required. In scenarios with $S_\mathrm{HP}>50\,\%$ the uncertainty of the savings increases as they exhibit both more and less cost-effective scenarios.
        Regarding the relative OPP, it can be observed that low shares of $25\,\%$ can actually be beneficial in lowering peak values. An effect that is especially pronounced in the suburban topology. In these scenarios, where the summer PV generation dominates the OPP values, the HPs represent additional flexibility that is used to lower OPP by shifting the drinking hot water production to more favorable hours. This flexibility could be further increased when using HPs that are also capable of cooling. However, once the share reaches $50\,\%$, the benefits start diminishing as the HP's demand in winter is responsible for most OPP occurrences. While the scenarios still show great variations at shares of $50\,\%$, the uncertainty diminishes while the median OPP value increases. In scenarios with $100\,\%$ HP penetration, the technical benefits of LEMs diminish as most OPP events occur in winter when no PV generation is available for several days, and thus, shifting of the load is not incentivized due to a lack of price signals.
        In summary, it is shown that HPs have significant positive economic impacts in LEMs albeit the marginal impact decreases when shares go beyond $50\,\%$. The inverse can be said about OPP values. While shares $<50\,\%$ generally have a positive impact as more flexibility is available during summer and transition periods, the impact becomes negative at higher shares as HPs become the dominant cause for peak loads.

\section{Discussion}\label{sec:discussion}

    \subsection{Implications of the findings on the energy system}\label{ssec:implications}
        This study provides a detailed examination of the techno-economic benefits and challenges associated with LEMs across various grid topologies. It highlights the potential of LEMs to contribute significantly to both economic efficiency and grid stability while also underscoring the limitations and specific conditions under which these benefits are realized. This discussion is intended to offer a comprehensive understanding of the economic and technical implications, suggest practical design considerations for LEMs, and identify the key limitations of our research methodology.
        
        \subsubsection{Economic implications}
            Our analysis indicates that LEMs can be economically viable, with a substantial majority (98\,\%) of scenarios showing reduced energy costs for participants. This demonstrates that even relatively simple market designs can effectively incentivize both producers and consumers to engage in energy trading within LEMs as both of them only stand to gain financially by participating in the market. This could also eradicate potential issues in regions where not all households want to participate in the LEM since everyone benefits. A notable potential benefit of implementing LEMs is the ability to restructure grid fees. Traditionally, grid fees have been assessed based on the voltage level at which a consumer is connected, assuming that lower voltage connections use more of the grid infrastructure \cite{StromNEV2005}. This assumption may no longer hold true where decentralized generation is common and occurs at mostly mid- and low-voltage levels. LEMs could facilitate more precise grid fee assessments by enabling the tracking of the origin of generation and destination of electricity consumption at the low-voltage level thus potentially lowering or more justly distributing costs depending on actual grid usage and encouraging local generation.
    
            However, it needs to be noted that the broader economic impact of LEMs, in particular the operational costs associated with managing these markets, was excluded in our study. As LEMs have only been implemented as pilots so far \cite{Mengelkamp2018, Luth2020, Ableitner2019}, a precise allocation of costs was not possible.
            Our investigations estimated cost savings of up to 20\,\% without the consideration of a restructured grid-fee system. All excluded costs need to sum up to less to make the concept economically viable overall. Future research should focus on the comprehensive cost-benefit analysis of LEMs to ensure that the economic advantages outweigh the operational and maintenance costs associated with these systems. 
            Another issue that needs to be addressed is the relatively low liquidity in LEMs in comparison to national wholesale markets. An issue especially prevalent in the countryside and rural topology due to its limited number of households.

        \subsubsection{Technical implications}
            The technical analysis of LEMs in this study has revealed that they hold potential for managing operational peak load demands effectively, however, only under certain conditions. Specifically, our findings demonstrate that LEMs are most effective in scenarios where energy generation and energy consumption align. For instance, on sunny days when PV systems are most productive, LEMs can capitalize on the high availability of local generation, given there is enough flexibility to harness most of the generation. The increased synchronization allows for the reduction of imported energy transmitted, thereby decreasing the overall load on the grid's infrastructure, in particular its transformer, and enhancing its resilience. The effect is even more noticeable during transition periods when both PV generation and heating demand coincide.
    
            However, their effectiveness tends to diminish when the demand for energy substantially exceeds local generation capacities. This is particularly evident in colder months or in regions with significant reliance on heat pumps. During these times, the local generation from PV systems is reduced due to shorter daylight hours and lower solar yield, while energy demand increases due to heating needs. The same phenomenon can be observed in regions with high shares of PV but little flexible consumption in the summer months when the high PV output cannot be utilized due to the lack of electricity demand. This is supported by the observation that among the 179 scenarios where the operational peak load was reduced by at least 10\% with the LEM, only 5 scenarios had the share of HPs exceeding that of PV, and even then, by only 25\%. Additionally, only 2 scenarios featured solely PV.
            
\begin{figure}
                \centering
                \begin{subfigure}[b]{0.45\textwidth}
                    \centering
                    \includegraphics[width=\textwidth]{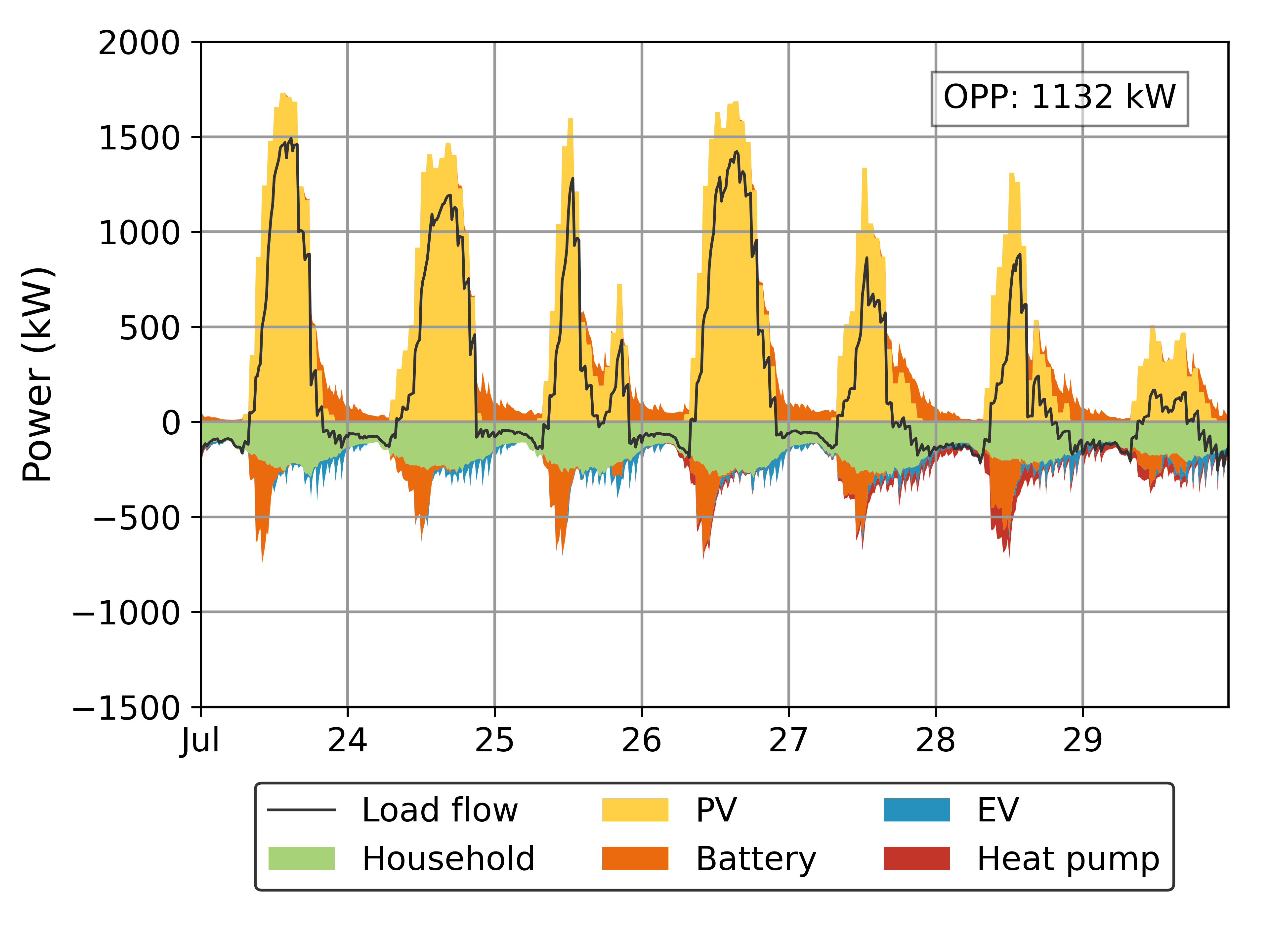}
                    \caption{Summer -- without LEM}
                    \label{fig:powerflow_summer_woLEM}
                \end{subfigure}
                \hfill
                \begin{subfigure}[b]{0.45\textwidth}
                    \centering
                    \includegraphics[width=\textwidth]{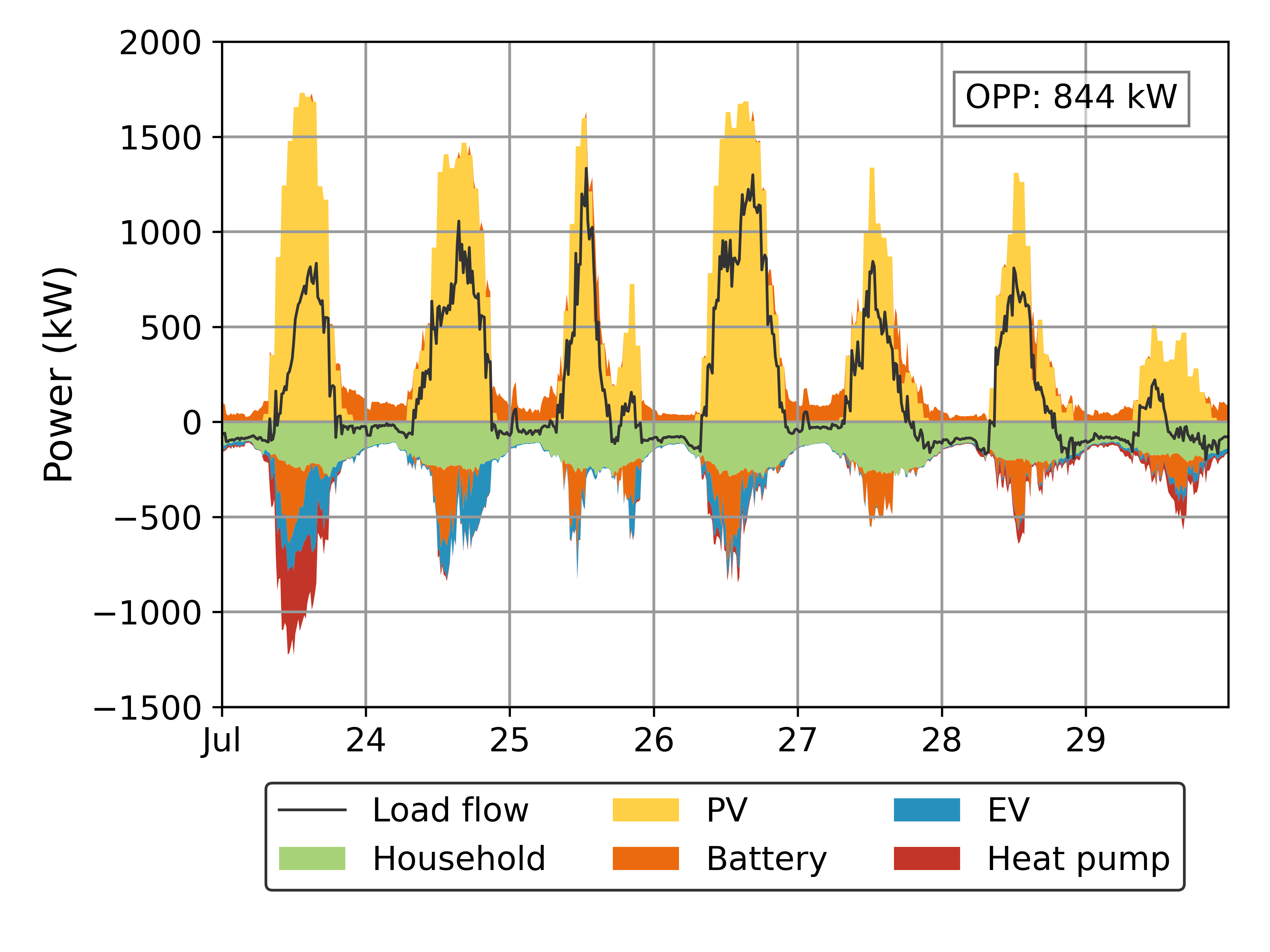}
                    \caption{Summer -- with LEM}
                    \label{fig:powerflow_summer_wLEM}
                \end{subfigure}
            
                \vspace{1em} 
                
                \begin{subfigure}[b]{0.45\textwidth}
                    \centering
                    \includegraphics[width=\textwidth]{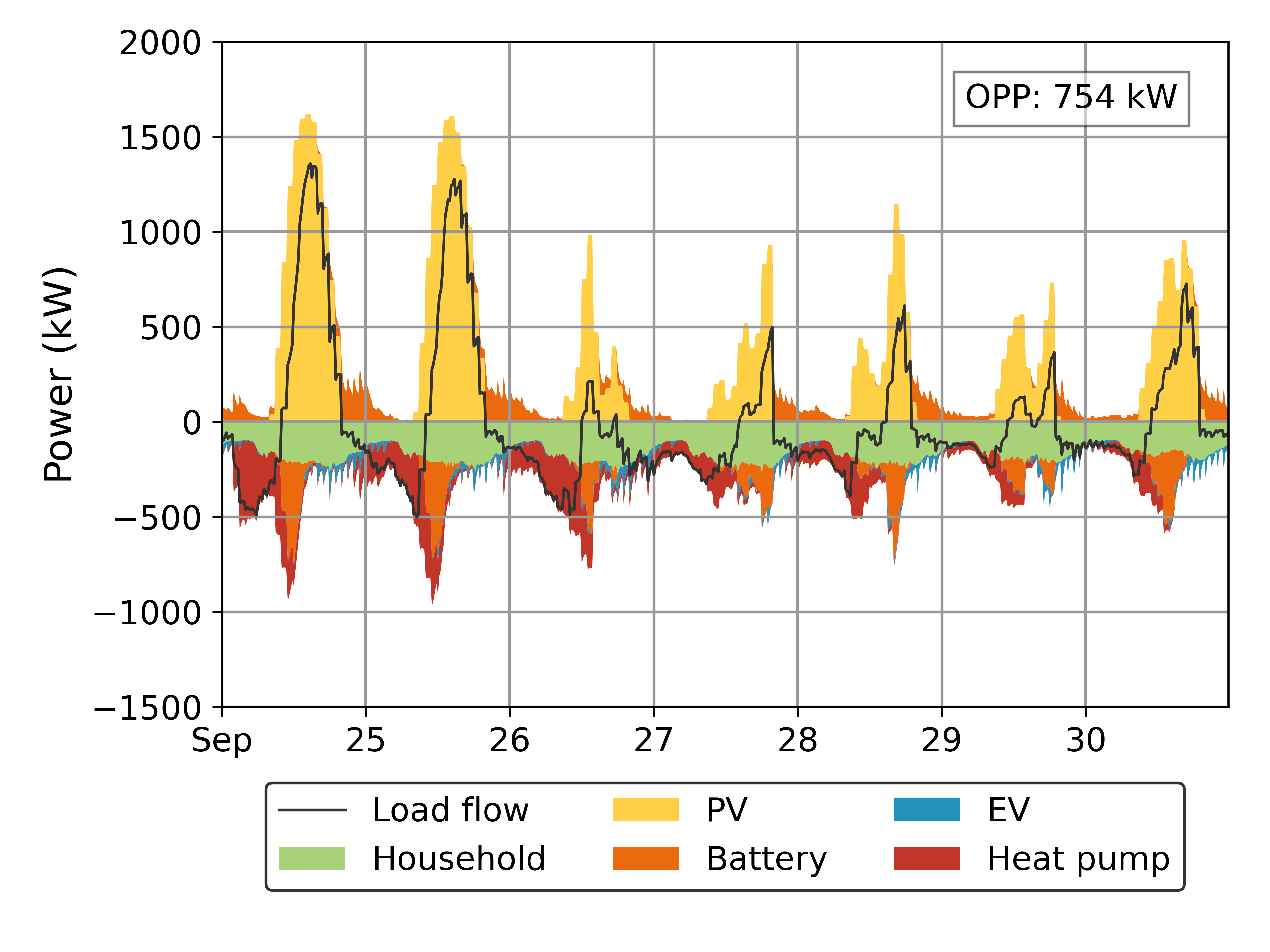}
                    \caption{Transition -- without LEM}
                    \label{fig:powerflow_transition_woLEM}
                \end{subfigure}
                \hfill
                \begin{subfigure}[b]{0.45\textwidth}
                    \centering
                    \includegraphics[width=\textwidth]{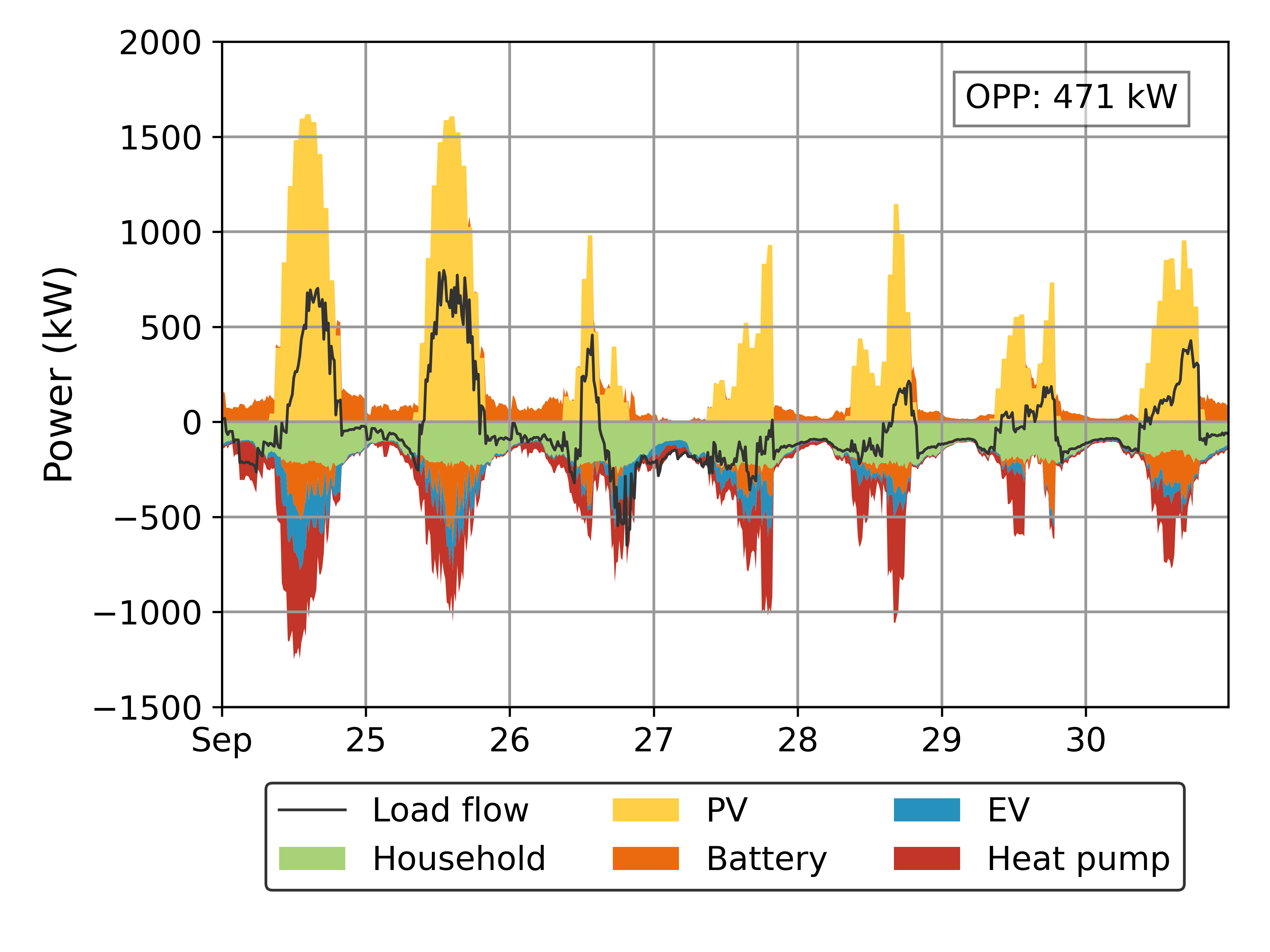}
                    \caption{Transition -- with LEM}
                    \label{fig:powerflow_transition_wLEM}
                \end{subfigure}
            
                \vspace{1em} 
                
                \begin{subfigure}[b]{0.45\textwidth}
                    \centering
                    \includegraphics[width=\textwidth]{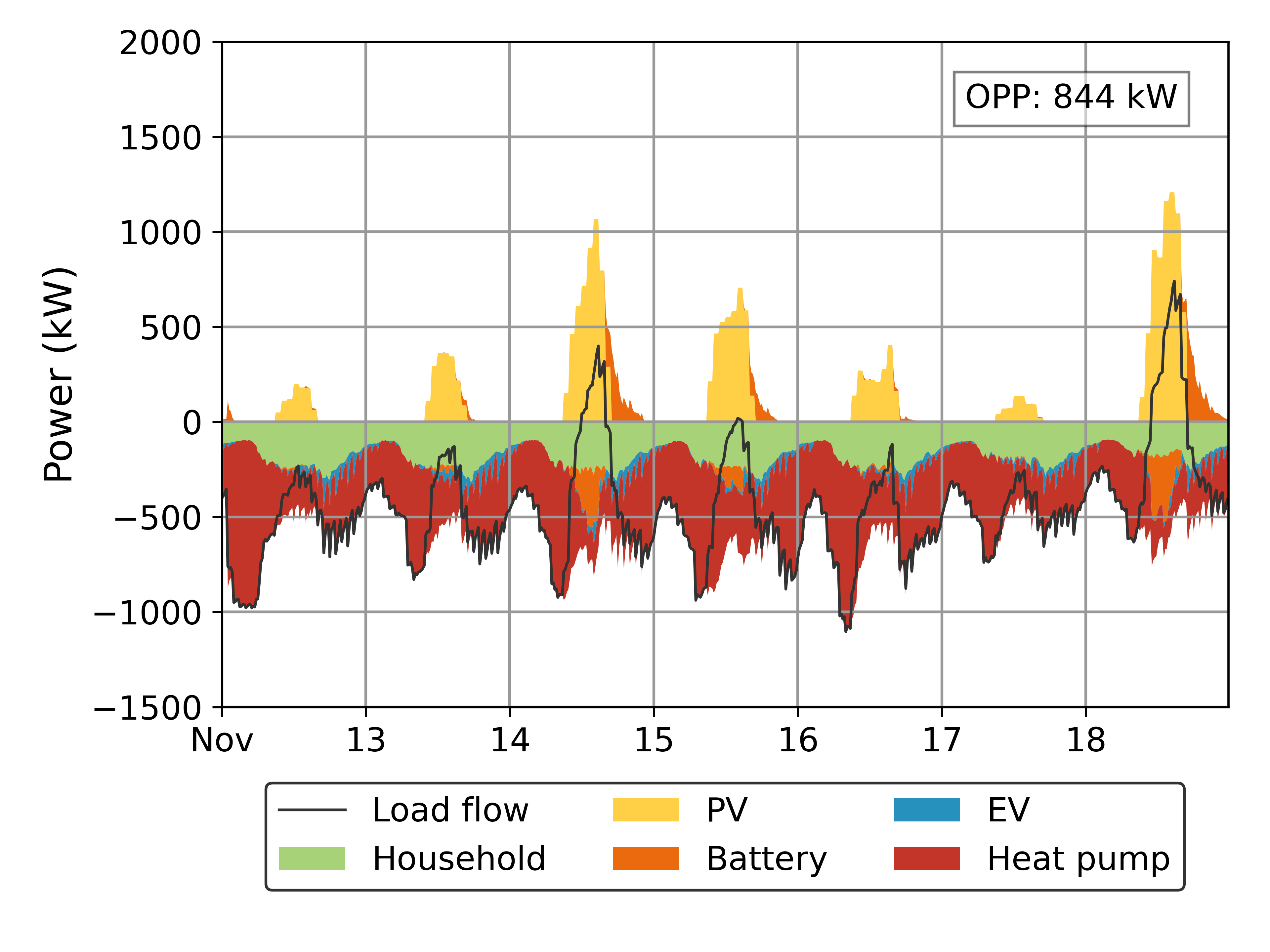}
                    \caption{Winter -- without LEM}
                    \label{fig:powerflow_winter_woLEM}
                \end{subfigure}
                \hfill
                \begin{subfigure}[b]{0.45\textwidth}
                    \centering
                    \includegraphics[width=\textwidth]{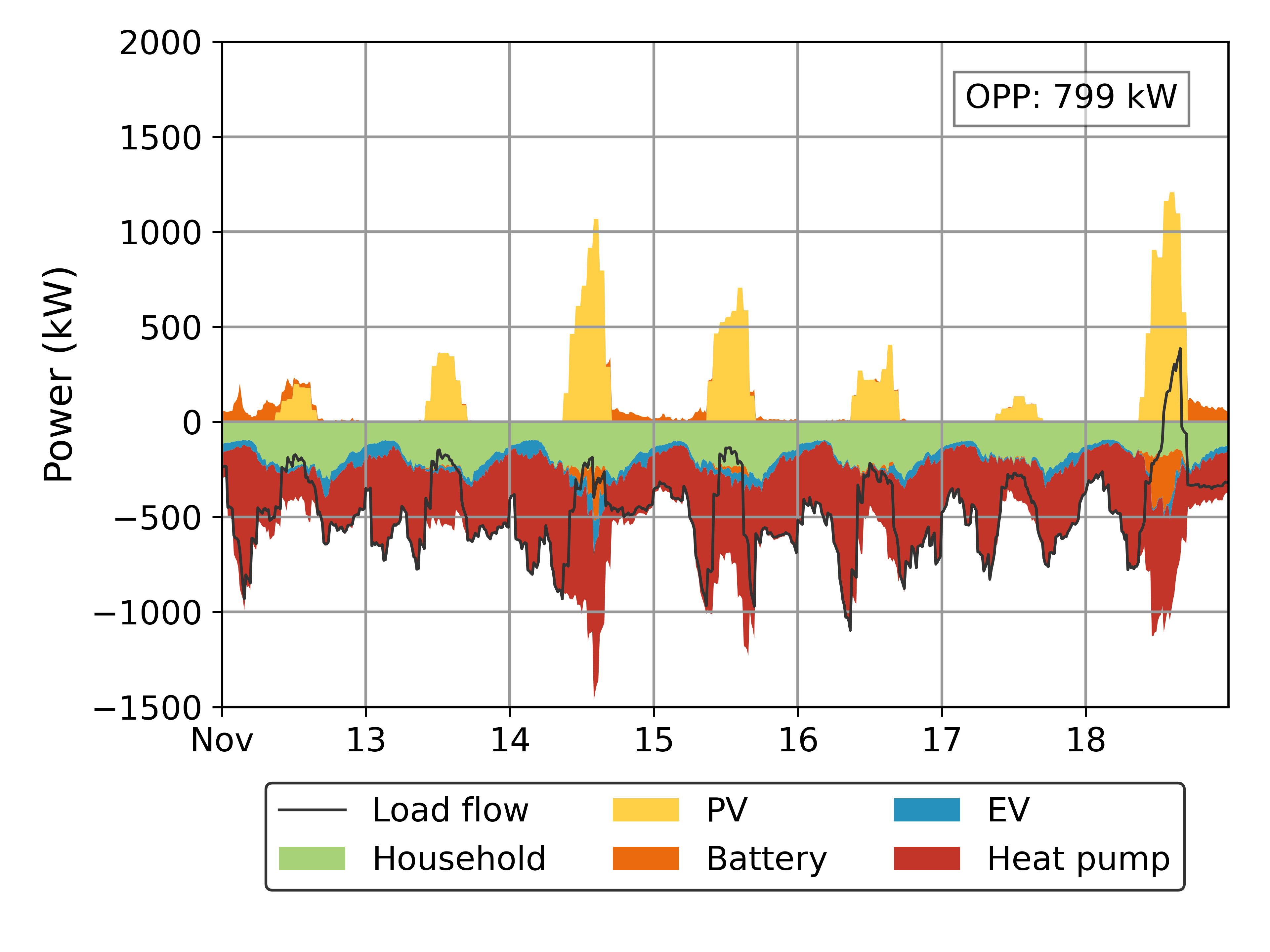}
                    \caption{Winter -- with LEM}
                    \label{fig:powerflow_winter_wLEM}
                \end{subfigure}
            
                \caption{Figure plotting the power flows of the system with and without LEM}
                \label{fig:powerflow}
            \end{figure}

            \autoref{fig:powerflow} in our analysis illustrates the aforementioned points. It compares the weekly power flows in an urban grid scenario with PV-, EV- and HP share at 100\,\%, both with and without a LEM, for all three representative weeks. Although the LEM manages to slightly reduce operational peak loads in winter, the reduction is modest—from 844 to 799\,kW, a decrease of merely 45\,kW. Contrast this with the summer and transition seasons depicted in the same figure, where reductions in OPP reach as high as 288 and 283\,kW, respectively. However, the operational peak loads reductions is drastically diminished without EVs and HP when the reductions in summer and transition diminish to 37 and 3\,kW, respectively (not shown in the figure).

            The observations underscore a critical design consideration for LEMs: the necessity to include mechanisms that can effectively mitigate peak loads even during low local generation periods. A possible strategy to reduce OPP consistently and reliably is to deploy more sophisticated market designs. For example, grid-aware clearing incorporates the actual grid into the clearing method and thus can deny trades beyond technical limitations \cite{Jodeiri-Seyedian2022, Majumdar2023, Gupta2023}. Another possibility is the deployment of redispatch measures in which the grid operator either indirectly influences controllable devices through economic incentives or directly through deactivation or reduction \cite{EnWG2024}.
            
            In conclusion, while LEMs offer promising avenues for enhancing grid stability and managing energy costs, their design must be carefully tailored to local conditions and include versatile mechanisms to handle the inherent variability in renewable energy generation and consumption patterns. This approach will ensure that LEMs not only support sustainable energy usage but also contribute to the resilience and reliability of the energy grid.
        
        \subsubsection{Implications for the design of LEMs}
            Our findings indicate several critical design principles for LEMs. First, the effectiveness of LEMs significantly increases with participant scale. Urban and suburban settings, with their dense populations, are particularly well-suited for LEM deployment. Interestingly, despite the smaller available area for PV installations per household in urban areas, these settings benefit technical aspects. Urban households typically demonstrate lower specific energy demands, effectively balancing the reduced energy generation with lower heating demands.
            Second, integrating photovoltaics within LEMs proves advantageous, enhancing local generation capacity to effectively manage and reduce peak loads. Across all grid topologies examined, increasing the share of PV consistently improves both economic and technical outcomes. The exceptions to this rule occur in scenarios lacking electric vehicles and heat pumps, where the benefits of PV are not fully realized.
            Conversely, the impact of electric vehicles on grid stability and costs is minimal, suggesting that widespread adoption of EVs is unlikely to pose significant challenges to grid operations as long as PV and/or HP are also present in the grid. 
            However, heat pumps present a complex dual role. Economically, they lower energy costs when their share is high, yet technically, they introduce challenges. Specifically, in scenarios where heat pump shares exceed 50\,\%, the benefits of LEMs diminish as winter peak loads increase unless there is an equivalent or greater share of PV systems to offset these loads.
            Lastly, our analysis of representative weeks indicates that LEMs are more beneficial in climates where demand and generation align more consistently. Unlike the stark distinction between generation in summer and demand in winter typical of Central European climates, the Southern European climate -- with its lower heating demand and higher cooling demand in summer -- presents more favorable conditions for LEMs. Conversely, our findings suggest that LEMs in more northern regions may offer fewer benefits due to their challenging climatic conditions.
            
    \subsection{Key limitations}
        While our approach is thorough, it encompasses several key limitations that influence the interpretation of our results. Due to the computationally intensive nature of agent-based models, we constrained our simulation scope. Instead of simulating an entire year, we opted for three representative weeks per season to depict typical operating conditions of the distribution grids. This method, while efficient, does not capture extreme conditions or anomalous events that could occur outside these periods, potentially omitting crucial data that could affect our understanding of grid dynamics under stress.
        Furthermore, our models used a single example grid for each grid topology. This approach does not account for the considerable variation in grid designs, especially in systems with few participants where unique elements such as schools or large commercial facilities can significantly alter grid behavior. We acknowledge that including a broader array of grid samples would enhance the reliability of our conclusions but would also increase computational demands substantially.
        Additionally, our study is based on grid topologies and weather data sourced from a German grid operator, limiting the generalizability of our findings. While the grid topologies are similar across Europe, the climate is a critical factor for both PV generation and energy demand, and our results may not directly translate to regions with different climatic conditions or building insulation standards.
        Lastly, the simulation employed a singular market and agent setup, focusing on a continuous market model. This setup does not explore how alternative models, such as day-ahead markets, might impact LEM performance, given that forecast accuracy plays a crucial role in their effectiveness. Exploring diverse market designs could potentially alter the economic and technical viability of the grids, but such an exhaustive investigation was beyond this study’s scope.

    \subsection{Conclusion}
        In this paper, we investigated the economic and technical influences of LEMs using 100 different combinations of PV, EV, and HP penetration levels in four different grids, totalling 400 different scenarios.
        We found that 99\,\% of the scenarios exhibit a lower average energy price, and 80\,\% exhibit a lower operational peak power.
        We observed that the share of PV and HP have the most profound impact on both the economic and technical scale while EVs have only a small influence. From our results we deduce that the technical hurdles for the integration of these technologies will be set by the shares of PV and HP regardless of the existence of an LEM or not. EVs on the other hand should be fairly easy to integrate due to the low timely correlation factor of their charging.
        From an economic perspective we observed that LEMs were most effective in scenarios with high shares of PV and HP with a somewhat smaller positive impact from high EV shares.
        From a technical perspective LEMs were able to relatively decrease the OOP the most in scenarios with high shares of PV yet lower shares of HPs or a high ratio of PV generation to heat demand.
        These observations were consistent across all topologies, although, they could be more reliably made for grids with more participants, i.e. the suburban and urban grid. 
        Overall, it was noticeable that LEMs are especially beneficial in periods when generation and (flexible) demand align such as in summer and especially the transitions phases in Central Europe. In winter, when there is very little generation, the economic incentives set by the LEM are too small for the demand to shift significantly thereby reducing the load on the grid.  We therefore recommend that other mechanisms are required to ensure safe operation of the grid especially in areas with small PV shares or in winter. This issue is most likely to be more prevalent further North where the disparity between winter PV generation and heating demand is far greater and less so in more Southern regions where both winter and summer generation and demand are more aligned.

\section{Experimental procedures}

    \subsection{Resource availability}
    
        \subsubsection{Lead contact}
            Further information and requests for resources and data should be directed to and will be fulfilled by the lead contact, Markus Doepfert (\url{markus.doepfert@tum.de})
            
        \subsubsection{Materials availability}
            This study did not generate new unique materials.
            
        \subsubsection{Data and code availability}
            The code is available at \url{https://github.com/TUM-Doepfert/lemlab/tree/doepfert2024_lem}. All scenarios and scenario results are available at \url{https://zenodo.org/records/13623929} (DOI: 10.5281/zenodo.13623929). A compact version of the results that only contains the relevant files for the analysis and figure creation for this paper can be found under \url{https://zenodo.org/records/13907329} (DOI: 10.5281/zenodo.13907329).

    \subsection{Methodology}\label{ssec:methodology}
    
        The methodology for this paper can be divided into two parts which will be explained subsequently. At first, the various components were sized optimally for each agent. Afterward, the models were run once with and once without a local energy market.
        
        \subsubsection{Optimal building component sizing}
            
            For the sizing of distributed energy resources (DERs) within each region, we utilized the optimization-based model framework \textit{urbs}, detailed in \cite{Candas2023}. This model employs mixed-integer linear programming (MILP) to determine the cost-optimal operation and dimensioning of DERs while maintaining the balance of electrical, heat, and mobility supply and demand. The DERs considered include solar PV, heat pumps, EVs (with home charging), batteries, and thermal storage systems. Unlike the market simulation step, which is described below, this dimensioning assumes perfect forecasts of the model time series.
            
        \subsubsection{Simulation of scenarios}
        
            The simulations run either with only the wholesale market to represent today's European market design or also with an LEM. The simulations were run in an adapted version of the tool \textit{lemlab} \cite{lemlab2023}.
        
            \paragraph{Market design}
            
                The wholesale energy trade in the European Union is an energy-only market and is split into several different markets with different timescales. Trades occur both over-the-counter as well as on dedicated markets such as the futures and spot markets \cite{Growitsch2009}. 
                The end consumer of energy barely notices the intricate interactions that occur behind the scenes. For them, the energy price remains constant regardless of when and how much energy is consumed in Germany. Therefore, wholesale trading was modeled accordingly as a constant source of energy for a constant price.
                
                The same is true for feeding in energy. Small-scale renewable energy producers receive a fixed feed-in tariff regardless of the amount and time they produce the energy for the first 20 years since installation \cite{EEG2014}. Thus, the energy production is compensated with a fixed value.
                
                The idea of local energy markets has long been discussed \cite{Collier1994, Kamrat2001, Lund2004}. In principle, they are small-scale markets that are limited to a small geographic area. Typically, they are used in low-voltage distribution grids, which are mostly characterized by residential participants. The participants are able to trade energy freely which is facilitated by an LEM \cite{Khorasany2018}. The aim of the LEMs is that through their trade participants are able to offer their surplus and cover their deficits within the low-voltage system without relying on the higher-level grid levels. In principle, this should lead to lower peak loads in upstream grid levels as well as lower energy costs due to efficiency gains. 
                
                The design of LEMs can vary widely. They can either be organized in a decentralized fashion, where participants trade directly with each other, or a centralized one, where participants are connected through a platform that ensures that supply and demand are in balance and the grid is stable. Another option is a hybrid design in which both types of trades are allowed \cite{Khorasany2018}. They can be either cooperative, where a central entity coordinates all participants, or competitive, where each participant tries to maximize their own gain \cite{Bjarghov2021}. The market can be cleared using distributed, optimization-based or auction-based methods as it already occurs in today’s spot markets. The LEM participants are usually linked to upper-level markets through a mediator which is typically the wholesale energy retailer but does not have to be \cite{Khorasany2018}. 
                
                In accordance with the spot markets designs this study uses a centralized and competitive design with a periodic double-sided auction. The market price is calculated as pay-as-cleared. Thus, every participant obtains the same price in an auction. The auctions occur with a 24-hour rolling horizon in a 15-minute interval. Therefore, the participants can trade 96 times for every point in time. This design was chosen as it represents the European goals for the future energy markets where trading occurs closer to the point of delivery and in a liberalized manner.
                
            \paragraph{Simulation tool}
                
                The simulations with and without LEM were conducted using \textit{lemlab} (\textbf{l}ocal \textbf{e}nergy \textbf{m}arket \textbf{lab}oratory) \cite{lemlab2023}. \textit{lemlab} is an open-source, agent-based tool for the simulation of LEMs and has been previously used for research, e.g. to asses the flexibility of heat pumps in LEMs \cite{You2024}. Each household is modeled individually thus allowing each participant to optimize their operation schedule. The following actions are performed by each participant at every time step (optional steps are skipped if the simulation runs without a LEM):
                
                \begin{enumerate}
                    \item Obtain previous market results (optional): Each participant obtains the market results of the last clearing time 
                    \item Set target grid power: The real-time controller adjusts generation and consumption based on the available power as well as the purchased and sold energy to ensure that the grid is balanced for the current time step
                    \item Update forecasts (optional): Update the forecasts for demand and generation as well as energy prices
                    \item Optimize operation schedule (optional): The model-predictive controller adjusts the operation schedule based on the updated forecasts, the bidding strategy and past market clearing results for future time steps
                    \item Post bids and offers (optional): The participants post their bids and offers to the LEM
                \end{enumerate}
                
\begin{figure}
                    \centering
                    \includegraphics[width=\textwidth]{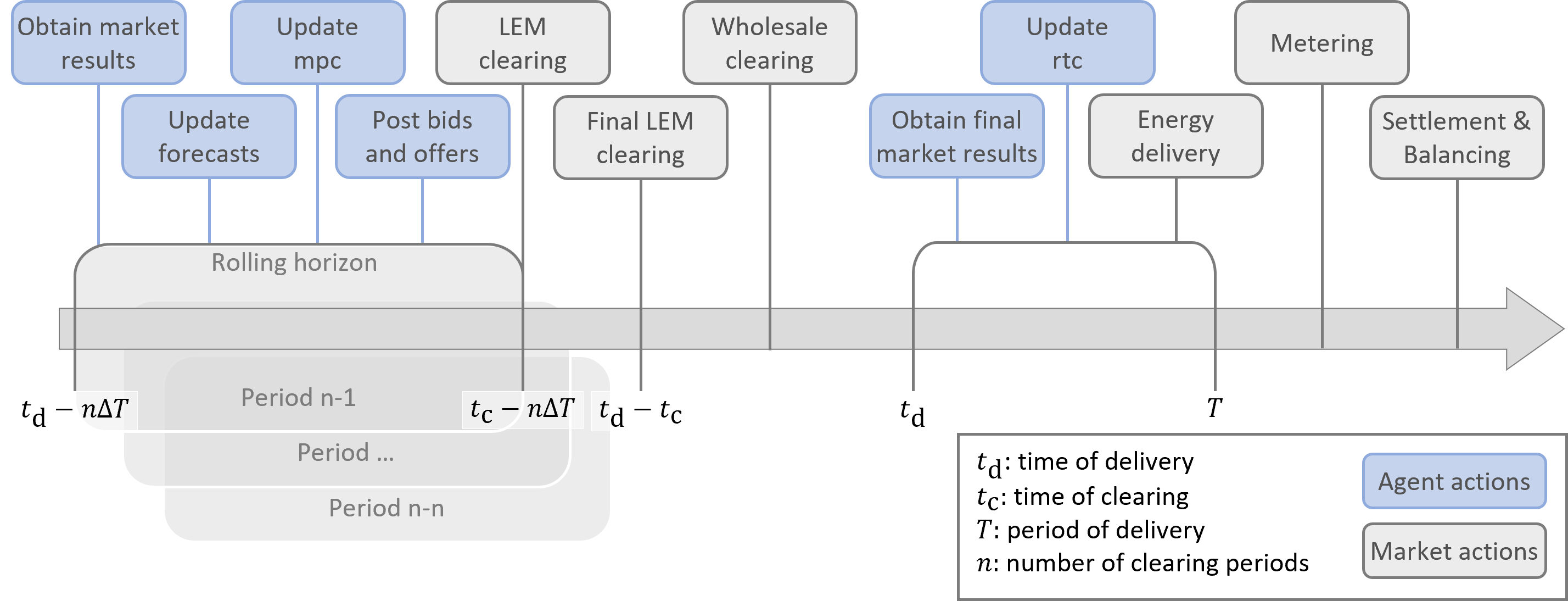}
                    \caption{The trading procedure for one time step in \textit{lemlab}}
                    \label{fig:lem_structure}
                \end{figure}

                After completing the cycle, bids and offers on the LEM are cleared. The cycle then repeats n times, allowing participants to adjust their operation schedules, bids, and offers. At the last time step before delivery, the final LEM clearing as well as the wholesale clearing occurs. Eventually, the energy is delivered, meters are updated and the market is settled.                
                This process is illustrated in \autoref{fig:lem_structure}. Participants unable to balance their supply and demand must pay a balancing fee to the mediator, reducing the likelihood of strategic bidding without incentive to maintain grid balance.

\subsection{Case study}
    
    To conduct the simulations, we first obtained input data for each agent (i.e., each building), derived representative grids for each topology, and set up realistic market parameters. The following subsections detail these steps.
    
    \subsubsection{Input Data}
        Agent-based simulations require extensive input data. Ideally, all data would be sourced from a single repository to ensure consistency and realistic household modeling. However, such comprehensive datasets are unavailable. Therefore, we combined real-life and synthetically generated data to create a reliable dataset, encompassing weather data, electricity demand, heating and mobility demand, and heat pump COPs. The resulting demands for each grid topology are shown in \autoref{tab:energy_demand}.
        
        \begin{table}[]
            \centering
            \caption{Demand characteristics of each grid region.}
            \label{tab:energy_demand}
            \begin{tabular}{rrrrr}
                \toprule
                              & \multicolumn{4}{c}{Annual energy demand {[}MWh{]}}      \\ \cline{2-5} 
                              & Countryside     & Rural    & Suburban    & Urban    \\ \hline
                Electricity          & 90              & 228        & 692         & 1581   \\
                Heat          & 875             & 1821       & 4866        & 6921   \\
                EV-charging (at home) & 13              & 52         & 179         & 212    \\ 
                \bottomrule
            \end{tabular}
        \end{table}
        
        \paragraph{Weather}
            Weather patterns significantly influence the optimal sizing of energy systems, making the use of representative weather data crucial. Lower ambient temperatures correlate with higher demands for space heating in buildings and affect heat pump performance, while solar irradiation levels determine the energy generation capacity of PV systems and contribute to solar heat gains in buildings. For this study, the Test Reference Year (TRY) data provided by the German Weather Service (DWD) was used, outlining hourly meteorological data representative of a typical year's weather in a given climate zone. This data, derived from weather records spanning 1995 to 2012, reflects current meteorological conditions~\cite{GermanWeatherServiceDWD2017}.
            
        \paragraph{Electric load}
            The electricity demand time series are 15-minute measurements that were taken from 01.05.2009 to 30.04.2010 in Bavaria, Germany for over 2,400 households~\cite{Vogt2011}. The households did not contain any heat pumps, air-conditioning, or electric vehicles, and therefore, their measurements contain the unaltered household demand. We subsequently matched the data to the time span of the weather data of 2015. The time series were mapped to the actual yearly demand of each household by choosing the time series with the closest yearly demand. Subsequently, we sized the original time series to the exact yearly demand of the household to ensure that both the time varying characteristics as well as the total yearly demand match.
        
        \paragraph{Heating load}
            The UrbanHeatPro tool was used to determine the heat demands in the considered buildings~\cite{Molar-Cruz2018}. This tool calculates the heat generation required to maintain comfort temperatures using an activity-based resistor-capacitor (RC) model, based on building characteristics such as use type, size, construction year, and refurbishment level, along with time series for solar and internal gains. 
            
            The building stock data for the considered regions was generated using a combination of open sources:
            
            \begin{itemize}
                \item LoD2 geometry data provided by the Bavarian surveying authority~\cite{BayerischeVermessungsverwaltung2024},
                \item the German census of 2011 for number of residents and construction years~\cite{StatistischeAemterdesBundesundderLaenderStatisticsOffices},
                \item TABULA building topology for the refurbishment ratios as well as the R and C values \cite{Loga2016}, and
                \item direct inspection of the area.
            \end{itemize}  
            
            The comfort temperature bands were sourced from various sources as they differ on the building types. The following building categories were differentiated:
            
            \begin{itemize}
                \item \textit{Residential buildings}: SIA 382/1 norm of the Swiss Society of Engineers and Architects on \textit{General Basics and Requirements of Ventilation and Air Conditioning Systems} as detailed in \cite{Dentel2006}.
                \item \textit{Public buildings}: Energy saving targets from the German Government, 2022 \cite{GermanGovernment2023}.
                \item \textit{Commercial and industrial buildings}: Requirements from the German Federal Ministry of Labour and Social Affairs for comfortable working conditions \cite{WorkplaceSafetyCommissionsoftheGermanFederalMinistryofLabourandSocialAffairs2022}.
            \end{itemize}

        \paragraph{PV generation}
            PV capacity factor time series were generated using the \textit{gsee} (Global Solar Energy Estimator) workflow, based on the aforementioned weather data~\cite{Pfenninger2016}. The capacity factors were calculated for each individual roof section of every building, reflecting expected higher energy yields for south-aligned modules.
        
        \paragraph{EV mobility}
            For simulating the electricity demands of the battery electric vehicles (BEVs), the  \textit{emobpy} tool is utilized~\cite{Gaete-Morales2021}. Using the mobility statistics from the \textit{Mobility in Germany} study \cite{Nobis2018}, an estimate on the car ownership and driving profiles (full- or part-time commuter, leisure driver) is made for each region (in an approach similar to \cite{Gaete-Morales2021}). Then, charging consumption profiles for a corresponding number of BEVs for each grid region is generated and assigned to each prosumer. In this, an \textit{immediate} charging strategy is assumed, i.e., the BEVs are allowed to charge wherever the possibility exists (including public or workplace charging). In turn, only the portion of  charging taking place at home is considered within the model's scope. 

        \subsubsection{Grid topologies}\label{ssec:grid_topology}
\begin{figure}
                \centering
                \begin{subfigure}{0.24\textwidth}
                    \centering
                    \includegraphics[width=\textwidth]{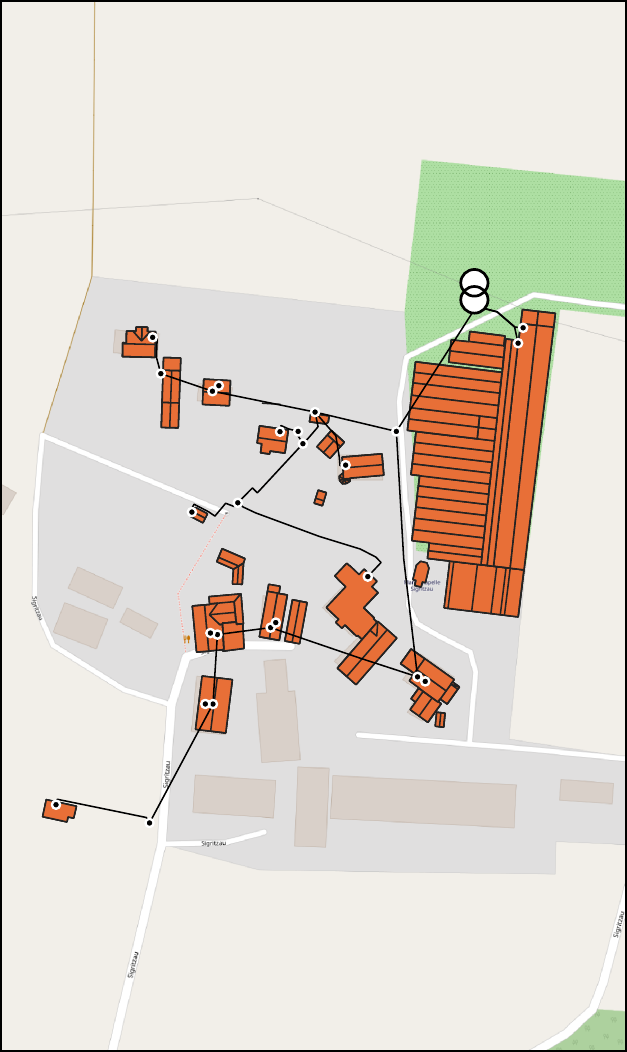}
                    \caption{Countryside}
                \end{subfigure}      
                \begin{subfigure}{0.24\textwidth}
                    \centering
                    \includegraphics[width=\textwidth]{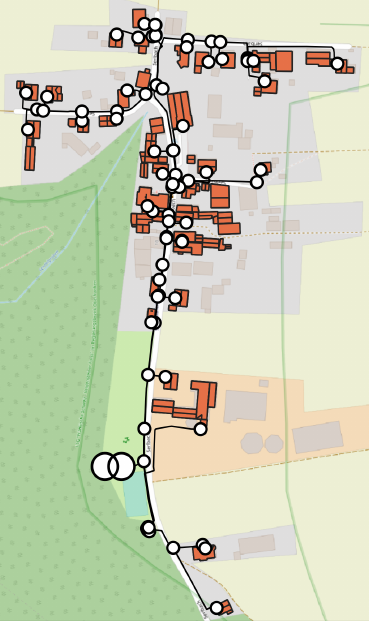}
                    \caption{Rural}
                \end{subfigure}      
                \begin{subfigure}{0.24\textwidth}
                    \centering
                    \includegraphics[width=\textwidth]{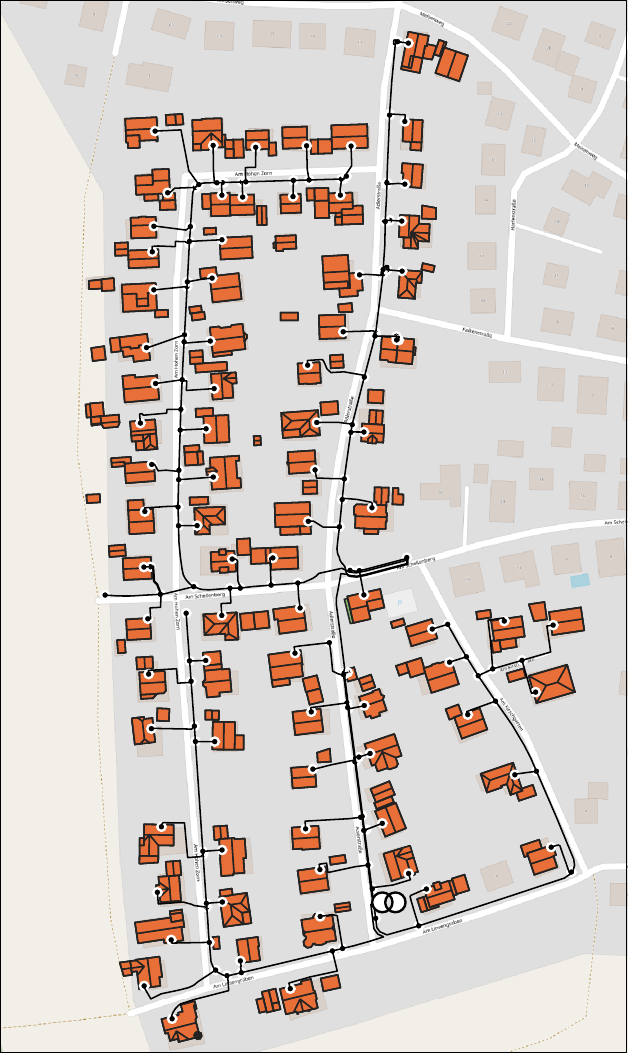}
                    \caption{Suburban}
                \end{subfigure}     
                \begin{subfigure}{0.24\textwidth}
                    \centering
                    \includegraphics[width=\textwidth]{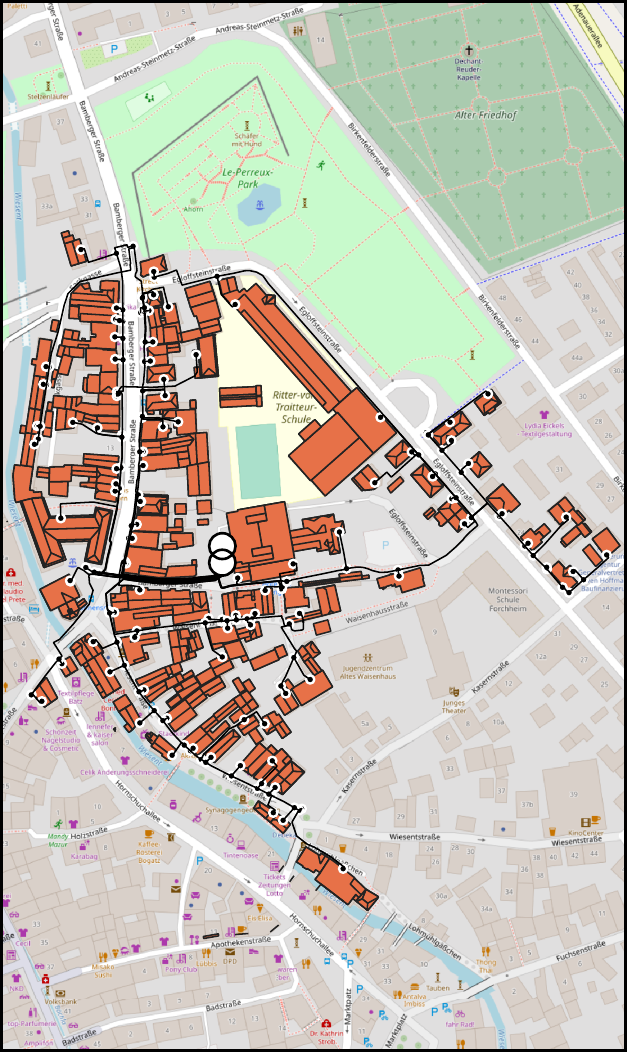}
                    \caption{Urban}
                \end{subfigure}
                
                \caption[The case study regions with the underlying grid topologies and the supplied buildings.]{The case study regions with the underlying grid topologies and the supplied buildings. Each building may consist of multiple roof sections, each of which is represented as an individual polygon in the maps \cite{Candas2024}.}
                \label{fig:grid_topologies}
            \end{figure}

            We considered four grid topologies to capture a wide range of real-world conditions (see \autoref{fig:grid_topologies}). These LV-grids are part of the node- and line-specific  distribution system modell (grid level 4-7) of a distribution system operator. Categorization of the LV-grids was done by using the methodology of Kerber looking on the number of loads, the mean line distance between the buildings and the ratio between residential and non-residential buildings~\cite{Kerber2008}:
            
            \begin{itemize}
                \item Countryside: Characterized by sparse populations and large distances between households.
                \item Rural: Moderate population density with a mix of residential and agricultural land use.
                \item Suburban: Higher population density with predominantly residential areas and some commercial activities.
                \item Urban: High population density with significant commercial activities.
            \end{itemize}

            All four LV-grids are radial systems and have specific grid structure data shown in \autoref{tab:grid_data}. The countryside grid has only single-family and duplex houses with mean line distances between the buildings greater than 50\,m. The rural grid is similar, but has a higher share of duplexes and some multi-family houses. Both grids feature a few non-residential buildings with agriculture purpose. The suburban grid has a mean line distance smaller than 30\,m and has a mix of residential buildings with high share of multi-family houses and only few commercial buildings. In contrast the urban grid has the comparatively highest share of commercials, such as schools, markets and restaurants. The mean line distance is smaller than 25\,m and most residential buildings are multi-family houses.
            
            \begin{table}[]
                \centering
                \caption{Grid structure data of the four LV-grids.}
                \label{tab:grid_data}
                \begin{tabular}{rrrrr}
                    \toprule
                                            & Countryside   & Rural     & Suburban    & Urban    \\ \hline
                    Transformer [kVA]       & 630           & 400       & 400       & 630       \\
                    Total cable length [m]  & 669           & 2,254     & 2,978     & 4,341     \\
                    Load buses              & 13            & 28        & 81        & 84        \\
                    Residential             & 13            & 57        & 186       & 378       \\
                    Non-residential         & 1             & 4         & 3           & 61        \\
                    \bottomrule
                \end{tabular}
            \end{table}
            
        \subsubsection{Scenario details} 
            The scenarios were designed to evaluate the performance of LEMs in four distinct grid topologies: countryside, rural, suburban, and urban. Each scenario varied the penetration levels of the key components PV, EVs and HPs. The penetration levels were incremented by 25\,\%, resulting in a comprehensive matrix of possible configurations. This subsection contains scenario details that have not been mentioned in the previous sections.
        
            \paragraph{HEMS parameters}
                The home energy management system plays a crucial role in the results of the simulations as it decides when energy is traded and at what price. It decides which forecasting methods to use for each component and how to plan the operation.
                
                The forecasts were set to "naive-average" which uses the average values of the previous two days for a time step to forecast the future value. The only component for which this does not work is the EVs. In their case, the so-called "ev-close" method was deployed. The HEMS has no information about the EV when it is not at home. However, when it returns, the HEMS is informed of how much energy was taken from the battery and when the EV will depart again. The operation was planned for the next 24 hours every 15 minutes.
                
                Market prices were forecasted using the "naive" forecasting using the value of the previous day. The trading horizon ranged between 3 and 24 hours. All agents used a linear trading strategy in which they linearly increased their bid price for purchased energy and decreased their ask price for sold energy until the threshold, i.e., the wholesale market prices, were reached 15 minutes before settlement.
                All parameters are listed in \autoref{tab:hems_params}.
                
                \begin{table}[h]
                    \centering
                    \caption{HEMS parameters}
                    \label{tab:hems_params}
                    \begin{tabular}{lr}
                        \toprule
                                                & \multicolumn{1}{c}{HEMS}   \\ \hline
                        Forecast electricity    & naive-average              \\
                        Forecast heat           & naive-average              \\
                        Forecast PV             & naive-average              \\
                        Forecast heat pump      & naive-average              \\
                        Forecast EV             & ev-close                   \\
                        Forecase market prices  & naive                      \\
                        Trading horizon {[}h{]} & 3-24                       \\
                        Trading strategy        & linear                     \\
                        \bottomrule
                    \end{tabular}
                \end{table}
            
            \paragraph{Market parameters}
                The market parameters were chosen to reflect the current German market design for the wholesale market. The consumer price comprises an energy (14.70\,ct/kWh) and a tax/levy (22.97\,ct/kWh) component. These were calculated using the average between 2020 and 2023, \cite{BDEW2023} to avoid choosing an outlier year. The same methodology was used to calculate the feed-in tariff, resulting in 8.27\,ct/kWh \cite{SFV2023}.
                
                Subsequently, the local energy market prices ranged between these two values. The market was designed as a periodic clearing market with a 24-hour horizon and 15-minute clearing intervals. Agents could trade up to 15 minutes before each time step. The clearing method was a double-sided auction, similar to European day-ahead markets. Agents unable to settle their energy needs incurred a balancing penalty of 1\,ct/kWh. Detailed market parameters are provided in \autoref{tab:market_params}.
                
                \begin{table}[h]
                    \centering
                    \caption{Market parameters for the wholesale and LEM}
                    \label{tab:market_params}
                    \begin{tabular}{lrr}
                        \toprule
                                                              & \multicolumn{1}{c}{Wholesale market} & \multicolumn{1}{c}{Local energy market} \\ \hline
                        Energy price {[}ct/kWh{]}             & 14.70                                & 8.27-14.70                              \\
                        Taxes \& levies {[}ct/kWh{]}          & 22.97                                & 22.97                                   \\
                        Feed-in tariff {[}ct/kWh{]}           & 8.27                                 & --                                      \\
                        Balancing price {[}ct/kWh{]}          & 14.70                                & 15.70                                   \\
                        Balancing feed-in tariff {[}ct/kWh{]} & 8.27                                 & 7.27                                    \\
                        Clearing horizon {[}h{]}              & --                                   & 24                                      \\
                        Clearing interval {[}min{]}           & --                                   & 15                                      \\
                        Clearing method                       & --                                   & Double-sided auction                    \\
                        \bottomrule
                    \end{tabular}
                \end{table}

\section{Acknowledgements}
    The authors thank the Bavarian Research Foundation ("Bayerische Forschungsstiftung"), Germany, as the presented work is funded by them through the research project \href{https://www.bayfor.org/de/unsere-netzwerke/bayerische-forschungsverbuende/forschungsverbuende/association/strom.html}{STROM}.
    
\section{Author contributions}
    Conceptualization, M.D. and S.C.;
    Methodology, M.D., S.C. and H.K.;
    Software, M.D., S.C. and H.K.;
    Formal Analysis, M.D.;
    Investigation, M.D.;
    Resources, T.H.;
    Data Curation, M.D.;
    Writing -- Original Draft, M.D., S.C. and H.K.;
    Writing -- Review \& Editing, M.D., S.C., H.K. and P.T.;
    Visualization, M.D.;
    Supervision, P.T. and T.H.;
    Project Administration, T.H.;
    Funding Acquisition, P.T. and T.H.

\section{Declaration of interests}
    The authors declare no competing interests.

\section{Declaration of generative AI and AI-assisted technologies in the writing process}
    Statement: During the preparation of this work the author(s) used ChatGPT in order to improve the quality of writing. After using this tool/service, the author(s) reviewed and edited the content as needed and take(s) full responsibility for the content of the published article.





\begin{thebibliography}{10}
\expandafter\ifx\csname url\endcsname\relax
  \def\url#1{\texttt{#1}}\fi
\expandafter\ifx\csname urlprefix\endcsname\relax\def\urlprefix{URL }\fi
\expandafter\ifx\csname href\endcsname\relax
  \def\href#1#2{#2} \def\path#1{#1}\fi

\bibitem{Gielen2019}
D.~Gielen, F.~Boshell, D.~Saygin, M.~D. Bazilian, N.~Wagner, R.~Gorini,
  \href{https://linkinghub.elsevier.com/retrieve/pii/S2211467X19300082}{The
  role of renewable energy in the global energy transformation}, Energy
  Strategy Reviews 24 (2019) 38--50.
\newblock \href {https://doi.org/10.1016/j.esr.2019.01.006}
  {\path{doi:10.1016/j.esr.2019.01.006}}.
\newline\urlprefix\url{https://linkinghub.elsevier.com/retrieve/pii/S2211467X19300082}

\bibitem{BDEW2023}
{Bundesverband der Energie- und Wasserwirtschaft (BDEW; German Association of
  Energy and Water Industries)},
  \href{https://www.bdew.de/media/documents/230215_BDEW-Strompreisanalyse_Februar_2023_15.02.2023.pdf}{{BDEW-Strompreisanalyse}}
  (Feb. 2023).
\newline\urlprefix\url{https://www.bdew.de/media/documents/230215_BDEW-Strompreisanalyse_Februar_2023_15.02.2023.pdf}

\bibitem{Needell2023}
Z.~Needell, W.~Wei, J.~E. Trancik,
  \href{https://linkinghub.elsevier.com/retrieve/pii/S2666386423000462}{Strategies
  for beneficial electric vehicle charging to reduce peak electricity demand
  and store solar energy}, Cell Reports Physical Science 4~(3) (2023) 101287.
\newblock \href {https://doi.org/10.1016/j.xcrp.2023.101287}
  {\path{doi:10.1016/j.xcrp.2023.101287}}.
\newline\urlprefix\url{https://linkinghub.elsevier.com/retrieve/pii/S2666386423000462}

\bibitem{Mugnini2023}
A.~Mugnini, F.~Polonara, A.~Arteconi,
  \href{https://www.sciencedirect.com/science/article/pii/S0378778823006461}{Quantification
  of the energy flexibility of residential building clusters: Impact of
  long-term refurbishment strategies of the italian building stock}, Energy and
  Buildings (2023) 113416\href {https://doi.org/10.1016/j.enbuild.2023.113416}
  {\path{doi:10.1016/j.enbuild.2023.113416}}.
\newline\urlprefix\url{https://www.sciencedirect.com/science/article/pii/S0378778823006461}

\bibitem{Loschan2023}
C.~Loschan, D.~Schwabeneder, G.~Lettner, H.~Auer,
  \href{https://linkinghub.elsevier.com/retrieve/pii/S0142061522007980}{Flexibility
  potential of aggregated electric vehicle fleets to reduce transmission
  congestions and redispatch needs: {{A}} case study in {{Austria}}},
  International Journal of Electrical Power \& Energy Systems 146 (2023)
  108802.
\newblock \href {https://doi.org/10.1016/j.ijepes.2022.108802}
  {\path{doi:10.1016/j.ijepes.2022.108802}}.
\newline\urlprefix\url{https://linkinghub.elsevier.com/retrieve/pii/S0142061522007980}

\bibitem{Qin2023}
X.~Qin, B.~Xu, I.~Lestas, Y.~Guo, H.~Sun,
  \href{https://linkinghub.elsevier.com/retrieve/pii/S2542435123002118}{The
  role of electricity market design for energy storage in cost-efficient
  decarbonization}, Joule 7~(6) (2023) 1227--1240.
\newblock \href {https://doi.org/10.1016/j.joule.2023.05.014}
  {\path{doi:10.1016/j.joule.2023.05.014}}.
\newline\urlprefix\url{https://linkinghub.elsevier.com/retrieve/pii/S2542435123002118}

\bibitem{EU2018}
\href{http://data.europa.eu/eli/dir/2018/2001/oj/eng}{Directive ({{EU}})
  2018/2001 of the {{European Parliament}} and of the {{Council}} of 11
  {{December}} 2018 on the promotion of the use of energy from renewable
  sources (recast) ({{Text}} with {{EEA}} relevance.)} (Dec. 2018).
\newline\urlprefix\url{http://data.europa.eu/eli/dir/2018/2001/oj/eng}

\bibitem{EU2019}
\href{http://data.europa.eu/eli/dir/2019/944/oj/eng}{Directive ({{EU}})
  2019/944 of the {{European Parliament}} and of the {{Council}} of 5 {{June}}
  2019 on common rules for the internal market for electricity and amending
  {{Directive}} 2012/27/{{EU}} (recast) ({{Text}} with {{EEA}} relevance.)}
  (Jun. 2019).
\newline\urlprefix\url{http://data.europa.eu/eli/dir/2019/944/oj/eng}

\bibitem{Khorasany2018}
M.~Khorasany, Y.~Mishra, G.~Ledwich, Market framework for local energy trading:
  A review of potential designs and market clearing approaches, Iet Generation
  Transmission \& Distribution 12~(22) (2018) 5899--5908.
\newblock \href {https://doi.org/10.1049/iet-gtd.2018.5309}
  {\path{doi:10.1049/iet-gtd.2018.5309}}.

\bibitem{Zaidi2018}
B.~H. Zaidi, S.~H. Hong,
  \href{https://doi.org/10.1007/s00202-017-0570-y}{Combinatorial double
  auctions for multiple microgrid trading}, Electrical Engineering 100~(2)
  (2018) 1069--1083.
\newblock \href {https://doi.org/10.1007/s00202-017-0570-y}
  {\path{doi:10.1007/s00202-017-0570-y}}.
\newline\urlprefix\url{https://doi.org/10.1007/s00202-017-0570-y}

\bibitem{Hussain2018}
A.~Hussain, V.-H. Bui, H.-M. Kim, A {{Resilient}} and {{Privacy-Preserving
  Energy Management Strategy}} for {{Networked Microgrids}}, IEEE Transactions
  on Smart Grid 9~(3) (2018) 2127--2139.
\newblock \href {https://doi.org/10.1109/TSG.2016.2607422}
  {\path{doi:10.1109/TSG.2016.2607422}}.

\bibitem{Gazafroudi2021}
A.~S. Gazafroudi, M.~Khorasany, R.~Razzaghi, H.~Laaksonen, M.~{Shafie-khah},
  \href{https://linkinghub.elsevier.com/retrieve/pii/S0306261921009521}{Hierarchical
  approach for coordinating energy and flexibility trading in local energy
  markets}, Applied Energy 302 (2021) 117575.
\newblock \href {https://doi.org/10.1016/j.apenergy.2021.117575}
  {\path{doi:10.1016/j.apenergy.2021.117575}}.
\newline\urlprefix\url{https://linkinghub.elsevier.com/retrieve/pii/S0306261921009521}

\bibitem{Ali2023}
L.~Ali, M.~I. Azim, J.~Peters, V.~Bhandari, A.~Menon, V.~Tiwari, J.~Green,
  S.~M. Muyeen, Application of a {{Community Battery-Integrated Microgrid}} in
  a {{Blockchain-Based Local Energy Market Accommodating P2P Trading}}, IEEE
  Access 11 (2023) 29635--29649.
\newblock \href {https://doi.org/10.1109/ACCESS.2023.3260253}
  {\path{doi:10.1109/ACCESS.2023.3260253}}.

\bibitem{Kim2023}
H.~J. Kim, Y.~S. Chung, S.~J. Kim, H.~T. Kim, Y.~G. Jin, Y.~T. Yoon,
  \href{https://www.sciencedirect.com/science/article/pii/S1364032123002927}{Pricing
  mechanisms for peer-to-peer energy trading: {{Towards}} an integrated
  understanding of energy and network service pricing mechanisms}, Renewable
  and Sustainable Energy Reviews 183 (2023) 113435.
\newblock \href {https://doi.org/10.1016/j.rser.2023.113435}
  {\path{doi:10.1016/j.rser.2023.113435}}.
\newline\urlprefix\url{https://www.sciencedirect.com/science/article/pii/S1364032123002927}

\bibitem{Bjarghov2021}
S.~Bjarghov, M.~L{\"o}schenbrand, A.~U.~N. Ibn~Saif, R.~A. Pedrero,
  C.~Pfeiffer, S.~K. Khadem, M.~Rabelhofer, F.~Revheim, H.~Farahmand,
  Developments and {{Challenges}} in {{Local Electricity Markets}}: {{A
  Comprehensive Review}}, IEEE Access 9 (2021) 58910--58943.
\newblock \href {https://doi.org/10.1109/access.2021.3071830}
  {\path{doi:10.1109/access.2021.3071830}}.

\bibitem{Mengelkamp2018}
E.~Mengelkamp, J.~G{\"a}rttner, K.~Rock, S.~Kessler, L.~Orsini, C.~Weinhardt,
  \href{https://www.sciencedirect.com/science/article/pii/S030626191730805X}{Designing
  microgrid energy markets: {{A}} case study: {{The Brooklyn Microgrid}}},
  Applied Energy 210 (2018) 870--880.
\newblock \href {https://doi.org/10.1016/j.apenergy.2017.06.054}
  {\path{doi:10.1016/j.apenergy.2017.06.054}}.
\newline\urlprefix\url{https://www.sciencedirect.com/science/article/pii/S030626191730805X}

\bibitem{Vasconcelos2019}
M.~Vasconcelos, W.~Cramer, C.~Schmitt, A.~Amthor, S.~Jessenberger, C.~Ziegler,
  A.~Armstorfer, F.~Heringer,
  \href{https://www.cired-repository.org/handle/20.500.12455/363}{The
  {{PEBBLES}} Project -- Enabling Blockchain Based Transactive Energy Trading
  of Energy \& Flexibility within a Regional Market}, AIM, 2019.
\newblock \href {https://doi.org/10.34890/591} {\path{doi:10.34890/591}}.
\newline\urlprefix\url{https://www.cired-repository.org/handle/20.500.12455/363}

\bibitem{Worner2019}
A.~W{\"o}rner, A.~Meeuw, L.~Ableitner, F.~Wortmann, S.~Schopfer, V.~Tiefenbeck,
  \href{https://energyinformatics.springeropen.com/articles/10.1186/s42162-019-0092-0}{Trading
  solar energy within the neighborhood: Field implementation of a
  blockchain-based electricity market}, Energy Informatics 2~(S1) (2019) 11.
\newblock \href {https://doi.org/10.1186/s42162-019-0092-0}
  {\path{doi:10.1186/s42162-019-0092-0}}.
\newline\urlprefix\url{https://energyinformatics.springeropen.com/articles/10.1186/s42162-019-0092-0}

\bibitem{GermanFederalStatisticsOffice2021}
{German Federal Statistics Office},
  \href{https://www.destatis.de/DE/Themen/Gesellschaft-Umwelt/Umwelt/UGR/private-haushalte/Tabellen/stromverbrauch-haushalte.html}{{Stromverbrauch
  der privaten Haushalte nach Haushaltsgr{\"o}{\ss}enklasse (Electricity
  consumption of private households by household size class)}} (2021).
\newline\urlprefix\url{https://www.destatis.de/DE/Themen/Gesellschaft-Umwelt/Umwelt/UGR/private-haushalte/Tabellen/stromverbrauch-haushalte.html}

\bibitem{Schlemminger2022}
M.~Schlemminger, T.~Ohrdes, E.~Schneider, M.~Knoop,
  \href{https://www.nature.com/articles/s41597-022-01156-1}{Dataset on
  electrical single-family house and heat pump load profiles in {{Germany}}},
  Scientific Data 9~(1) (2022) 56.
\newblock \href {https://doi.org/10.1038/s41597-022-01156-1}
  {\path{doi:10.1038/s41597-022-01156-1}}.
\newline\urlprefix\url{https://www.nature.com/articles/s41597-022-01156-1}

\bibitem{StromNEV2005}
{Federal Government},
  \href{https://www.gesetze-im-internet.de/stromnev/}{{Stromnetzentgeltverordnung
  (Electricity Grid Fee Act)}} (2005).
\newline\urlprefix\url{https://www.gesetze-im-internet.de/stromnev/}

\bibitem{Luth2020}
A.~L{\"u}th, J.~Weibezahn, J.~M. Zepter,
  \href{https://www.mdpi.com/1996-1073/13/8/1993}{On {{Distributional Effects}}
  in {{Local Electricity Market Designs}}---{{Evidence}} from a {{German Case
  Study}}}, Energies 13~(8) (2020) 1993.
\newblock \href {https://doi.org/10.3390/en13081993}
  {\path{doi:10.3390/en13081993}}.
\newline\urlprefix\url{https://www.mdpi.com/1996-1073/13/8/1993}

\bibitem{Ableitner2019}
L.~Ableitner, A.~Meeuw, S.~Schopfer, V.~Tiefenbeck, F.~Wortmann, A.~W{\"o}rner,
  \href{http://arxiv.org/abs/1905.07242}{Quartierstrom -- {{Implementation}} of
  a real world prosumer centric local energy market in {{Walenstadt}},
  {{Switzerland}}} (Jul. 2019).
\newblock \href {http://arxiv.org/abs/1905.07242} {\path{arXiv:1905.07242}}.
\newline\urlprefix\url{http://arxiv.org/abs/1905.07242}

\bibitem{Jodeiri-Seyedian2022}
S.-S. {Jodeiri-Seyedian}, A.~Fakour, M.~Jalali, K.~Zare,
  B.~{Mohammadi-Ivatloo}, S.~Tohidi,
  \href{https://linkinghub.elsevier.com/retrieve/pii/S2352467722001795}{Grid-aware
  pricing scheme in future distribution systems based on real-time power
  tracing and bi-level optimization}, Sustainable Energy, Grids and Networks 32
  (2022) 100934.
\newblock \href {https://doi.org/10.1016/j.segan.2022.100934}
  {\path{doi:10.1016/j.segan.2022.100934}}.
\newline\urlprefix\url{https://linkinghub.elsevier.com/retrieve/pii/S2352467722001795}

\bibitem{Majumdar2023}
A.~Majumdar, O.~Alizadeh-Mousavi,
  \href{https://ietresearch.onlinelibrary.wiley.com/doi/10.1049/gtd2.12686}{Grid-aware
  provision and activation of fast and slow power flexibilities from
  distributed resources in low and medium voltage grids}, IET Generation,
  Transmission \& Distribution 17~(13) (2023) 2926--2937.
\newblock \href {https://doi.org/10.1049/gtd2.12686}
  {\path{doi:10.1049/gtd2.12686}}.
\newline\urlprefix\url{https://ietresearch.onlinelibrary.wiley.com/doi/10.1049/gtd2.12686}

\bibitem{Gupta2023}
R.~K. Gupta, \href{http://infoscience.epfl.ch/record/299705}{Methods for
  {{Grid-aware Operation}} and {{Planning}} of {{Active Distribution
  Networks}}}, Ph.D. thesis, {\'E}cole Polytechnique F{\'e}d{\'e}rale de
  Lausanne (EPFL), Lausanne (Jan. 2023).
\newline\urlprefix\url{http://infoscience.epfl.ch/record/299705}

\bibitem{EnWG2024}
{Federal Government},
  \href{https://www.gesetze-im-internet.de/enwg_2005/EnWG.pdf}{{Gesetz {\"u}ber
  die Elektrizit{\"a}ts- und Gasversorgung (Law for the Electricity and Natural
  Gas Supply)}} (2005).
\newline\urlprefix\url{https://www.gesetze-im-internet.de/enwg_2005/EnWG.pdf}

\bibitem{Candas2023}
S.~Candas, B.~Reveron~Baecker, A.~Mohapatra, T.~Hamacher,
  \href{https://www.sciencedirect.com/science/article/pii/S0306261923005111}{Optimization-based
  framework for low-voltage grid reinforcement assessment under various levels
  of flexibility and coordination}, Applied Energy 343 (2023) 121147.
\newblock \href {https://doi.org/10.1016/j.apenergy.2023.121147}
  {\path{doi:10.1016/j.apenergy.2023.121147}}.
\newline\urlprefix\url{https://www.sciencedirect.com/science/article/pii/S0306261923005111}

\bibitem{lemlab2023}
S.~Lumpp, M.~Zade, M.~Doepfert, Y.~Zhengjie,
  \href{https://github.com/tum-ewk/lemlab}{Lemlab}, tum-ewk (Jul. 2023).
\newline\urlprefix\url{https://github.com/tum-ewk/lemlab}

\bibitem{Growitsch2009}
C.~Growitsch, R.~Nepal,
  \href{https://onlinelibrary.wiley.com/doi/10.1002/etep.324}{Efficiency of the
  {{German}} electricity wholesale market}, European Transactions on Electrical
  Power 19~(4) (2009) 553--568.
\newblock \href {https://doi.org/10.1002/etep.324}
  {\path{doi:10.1002/etep.324}}.
\newline\urlprefix\url{https://onlinelibrary.wiley.com/doi/10.1002/etep.324}

\bibitem{EEG2014}
{Federal Ministry for Economic Affairs and Climate Action},
  \href{https://www.gesetze-im-internet.de/eeg_2014/}{{Gesetz f{\"u}r den
  Ausbau erneuerbarer Energien (German Renewable Energy Sources Act)}} (2014).
\newline\urlprefix\url{https://www.gesetze-im-internet.de/eeg_2014/}

\bibitem{Collier1994}
U.~Collier, \href{https://doi.org/10.1177/0958305X9400500402}{Local {{Energy
  Concepts}} in {{Germany}} -- {{An Environmental Alternative}} to
  {{Liberalisation}}?}, Energy \& Environment 5~(4) (1994) 305--326.
\newblock \href {https://doi.org/10.1177/0958305X9400500402}
  {\path{doi:10.1177/0958305X9400500402}}.
\newline\urlprefix\url{https://doi.org/10.1177/0958305X9400500402}

\bibitem{Kamrat2001}
W.~Kamrat, \href{http://ieeexplore.ieee.org/document/917583/}{Modeling the
  structure of local energy markets}, IEEE Computer Applications in Power
  14~(2) (2001) 30--35.
\newblock \href {https://doi.org/10.1109/67.917583}
  {\path{doi:10.1109/67.917583}}.
\newline\urlprefix\url{http://ieeexplore.ieee.org/document/917583/}

\bibitem{Lund2004}
H.~Lund, P.~A. Oestergaard, A.~N. Andersen, F.~Hvelplund, H.~Maeng,
  E.~Muenster, N.~I. Meyer,
  \href{https://www.osti.gov/etdeweb/biblio/20476680}{{Local energy markets;
  Lokale energimarkeder}}, {Technical Report} NEI-DK-4221, {U.S. Department of
  Energy Office of Scientific and Technical Information}, Denmark (Jan. 2004).
\newline\urlprefix\url{https://www.osti.gov/etdeweb/biblio/20476680}

\bibitem{You2024}
Z.~You, S.~D. Lumpp, M.~Doepfert, P.~Tzscheutschler, C.~Goebel,
  \href{https://linkinghub.elsevier.com/retrieve/pii/S0306261923016331}{Leveraging
  flexibility of residential heat pumps through local energy markets}, Applied
  Energy 355 (2024) 122269.
\newblock \href {https://doi.org/10.1016/j.apenergy.2023.122269}
  {\path{doi:10.1016/j.apenergy.2023.122269}}.
\newline\urlprefix\url{https://linkinghub.elsevier.com/retrieve/pii/S0306261923016331}

\bibitem{GermanWeatherServiceDWD2017}
{German Weather Service (DWD)},
  \href{https://www.bbsr.bund.de/BBSR/DE/forschung/programme/zb/Auftragsforschung/5EnergieKlimaBauen/2013/testreferenzjahre/try-handbuch.pdf}{{Locally
  accurate test reference years from Germany for average, extreme and future
  weather conditions (de: Ortsgenaue Testreferenzjahre von Deutschland f{\"u}r
  mittlere, extreme und zuk{\"u}nftige Wetterverh{\"a}ltnisse)}}, Tech. rep.,
  DWD (2017).
\newline\urlprefix\url{https://www.bbsr.bund.de/BBSR/DE/forschung/programme/zb/Auftragsforschung/5EnergieKlimaBauen/2013/testreferenzjahre/try-handbuch.pdf}

\bibitem{Vogt2011}
C.~Vogt, {Analyse des Einfluss von Smart-Metering auf den Strombedarf von
  privaten Haushalten [Analysis of Smart-Meters on the Power Consumption of
  Private Households]}, {Diploma thesis}, Technical University of Munich,
  Munich (2011).

\bibitem{Molar-Cruz2018}
A.~{Molar-Cruz}, L.~Odersky, M.~Doepfert, V.~Meier,
  \href{https://github.com/tum-ens/UrbanHeatPro}{{{UrbanEnergyPro}}} (2018).
\newline\urlprefix\url{https://github.com/tum-ens/UrbanHeatPro}

\bibitem{BayerischeVermessungsverwaltung2024}
{Bayerische Vermessungsverwaltung (Bavarian Survey Office)},
  \href{https://geodaten.bayern.de/opengeodata/OpenDataDetail.html?pn=lod2}{{{3D}}
  building models ({{LoD2}})}.
\newline\urlprefix\url{https://geodaten.bayern.de/opengeodata/OpenDataDetail.html?pn=lod2}

\bibitem{StatistischeAemterdesBundesundderLaenderStatisticsOffices}
{Statistische {\"A}mter des Bundes und der L{\"a}nder (Statistics Offices)},
  \href{https://atlas.zensus2011.de/}{{Zensus 2011 (Census 2011)}}.
\newline\urlprefix\url{https://atlas.zensus2011.de/}

\bibitem{Loga2016}
T.~Loga, B.~Stein, N.~Diefenbach,
  \href{https://linkinghub.elsevier.com/retrieve/pii/S0378778816305837}{{{TABULA}}
  building typologies in 20 {{European}} countries---{{Making}} energy-related
  features of residential building stocks comparable}, Energy and Buildings 132
  (2016) 4--12.
\newblock \href {https://doi.org/10.1016/j.enbuild.2016.06.094}
  {\path{doi:10.1016/j.enbuild.2016.06.094}}.
\newline\urlprefix\url{https://linkinghub.elsevier.com/retrieve/pii/S0378778816305837}

\bibitem{Dentel2006}
A.~Dentel, U.~Dietrich,
  \href{https://rom-umwelt-stiftung.de/wp-content/uploads/2006/02/Dokumentation_Thermische_Behaglichkeit.pdf}{Thermal
  comfort in buildings ({{Thermische Behaglichkeit}} - {{Komfort}} in
  {{Geb{\"a}uden}})}, Tech. rep., HafenCity Universit{\"a}t Hamburg (2006).
\newline\urlprefix\url{https://rom-umwelt-stiftung.de/wp-content/uploads/2006/02/Dokumentation_Thermische_Behaglichkeit.pdf}

\bibitem{GermanGovernment2023}
{German Government},
  \href{https://www.bundesregierung.de/breg-de/schwerpunkte/klimaschutz/energiesparmassnahmen-2078224}{{Ma{\ss}nahmen
  zum Energiesparen (Measures to save energy)}} (Feb. 2023).
\newline\urlprefix\url{https://www.bundesregierung.de/breg-de/schwerpunkte/klimaschutz/energiesparmassnahmen-2078224}

\bibitem{WorkplaceSafetyCommissionsoftheGermanFederalMinistryofLabourandSocialAffairs2022}
{Workplace Safety Commissions of the German Federal Ministry of Labour and
  Social Affairs},
  \href{https://www.baua.de/DE/Angebote/Regelwerk/ASR/ASR-A3-5}{{Technische
  Regeln f{\"u}r Arbeitsst{\"a}tten: Raumtemperatur (Technical Rules for
  Workplaces: Room Temperature)}} (2022).
\newline\urlprefix\url{https://www.baua.de/DE/Angebote/Regelwerk/ASR/ASR-A3-5}

\bibitem{Pfenninger2016}
S.~Pfenninger, I.~Staffell,
  \href{https://linkinghub.elsevier.com/retrieve/pii/S0360544216311744}{Long-term
  patterns of {{European PV}} output using 30 years of validated hourly
  reanalysis and satellite data}, Energy 114 (2016) 1251--1265.
\newblock \href {https://doi.org/10.1016/j.energy.2016.08.060}
  {\path{doi:10.1016/j.energy.2016.08.060}}.
\newline\urlprefix\url{https://linkinghub.elsevier.com/retrieve/pii/S0360544216311744}

\bibitem{Gaete-Morales2021}
C.~{Gaete-Morales}, H.~Kramer, W.-P. Schill, A.~Zerrahn,
  \href{https://www.nature.com/articles/s41597-021-00932-9}{An open tool for
  creating battery-electric vehicle time series from empirical data, emobpy},
  Scientific Data 8~(1) (2021) 152.
\newblock \href {https://doi.org/10.1038/s41597-021-00932-9}
  {\path{doi:10.1038/s41597-021-00932-9}}.
\newline\urlprefix\url{https://www.nature.com/articles/s41597-021-00932-9}

\bibitem{Nobis2018}
C.~Nobis, T.~Kuhnimhof,
  \href{https://www.mobilitaet-in-deutschland.de/archive/pdf/MiD2017_Ergebnisbericht.pdf}{Mobility
  in {{Germany}} - {{MiD}} result report ({{Mobilit{\"a}t}} in {{Deutschland}}
  - {{MiD Ergebnisbericht}}}, Tech. rep., {German Federal Ministry for Digital
  and Transport} (2018).
\newline\urlprefix\url{https://www.mobilitaet-in-deutschland.de/archive/pdf/MiD2017_Ergebnisbericht.pdf}

\bibitem{Kerber2008}
G.~Kerber, R.~Witzmann,
  \href{https://mediatum.ub.tum.de/doc/681082/681082.pdf}{{Statistische Analyse
  von NS-Verteilungsnetzen und Modellierung von Referenznetzen (Statistical
  Distribution Grid Analysis and Reference Network Generation)}}, ew - Magazin
  f{\"u}r die Energiewirtschaft (Magazine for the energy industry) 6 (2008)
  22--26.
\newline\urlprefix\url{https://mediatum.ub.tum.de/doc/681082/681082.pdf}

\bibitem{Candas2024}
S.~Candas, Optimization and data acquisition framework for low-voltage
  distribution systems in transformation, Ph.D. thesis, TU Munich, Munich (Aug.
  2024).

\bibitem{SFV2023}
{Solarenergie F{\"o}rderverein (SFV; Solar Power Association)},
  \href{https://www.sfv.de/lokal/mails/sj/verguetu}{{Historische
  Solarstrom-Verg{\"u}tungen im {\"U}berblick}} (2023).
\newline\urlprefix\url{https://www.sfv.de/lokal/mails/sj/verguetu}

\end{thebibliography}


\end{document}